# Modeling Earthen Dike Stability: Sensitivity Analysis and Automatic Calibration of Diffusivities Based on Live Sensor Data

N.B. Melnikova[a,b]*, V.V. Krzhizhanovskaya[a,b], P.M.A. Sloot [a,b] *

[a] *University of Amsterdam, The Netherlands*
[b] *National Research University ITMO, Russia*

The paper describes concept and implementation details of integrating a finite element module for dike stability analysis "*Virtual Dike*" into an early warning system for flood protection. The module operates in real-time mode and includes fluid and structural sub-models for simulation of porous flow through the dike and for dike stability analysis. Real-time measurements obtained from pore pressure sensors are fed into the simulation module, to be compared with simulated pore pressure dynamics. Implementation of the module has been performed for a real-world test case – an earthen levee protecting a sea-port in Groningen, the Netherlands. Sensitivity analysis and calibration of diffusivities have been performed for tidal fluctuations. An algorithm for automatic diffusivities calibration for a heterogeneous dike is proposed and studied. Analytical solutions describing tidal propagation in one-dimensional saturated aquifer are employed in the algorithm to generate initial estimates of diffusivities.

Keywords: dike stability, porous flow, diffusivity calibration, sensitivity analysis, live sensor data

## 1 Introduction

Regular floods pose a serious threat to human life, valuable property and city infrastructure. Many international projects are aimed at the development of flood protection systems [1], [2]. The EU FP7 project SSG4Env is focused on development of semantic sensor grids for environmental protection. Flood Probe (also funded by FP7) coordinates related work on combining sensor measurement techniques. Flood Control 2015 (NL) aims to share sensor measurements datasets and to provide a user interface to explore sensor data for researchers, technical maintainers and civil population. The IJkDijk project (NL) [3] is a project on experimental physical study of dike failure mechanisms. The tests are carried out on full-scale experimental dikes equipped with large sets of sensors. The project has produced extremely detailed datasets of sensor data, including pore pressures, inclinations, stresses and strains. The *UrbanFlood* EC FP7 project [4] unites the work on monitoring dikes with sensor techniques [6], physical study of dike failure mechanisms [1], and software development for dike stability analysis [8], [9], simulation of dike breaching, flood, and city evacuation [10], [11], [12].
The early warning system is a multi-component system that runs in a real-time mode, gathering and analyzing measurements form sensors installed in the dikes, predicting dike stability, possibility of flooding and optimal evacuation routes. General workflow and interaction of software components in the *UrbanFlood* early warning system are presented in Figure 1. The *Sensor Monitoring* module receives data streams from the sensors installed in the dike. Raw sensor data is filtered by the *AI (Artificial Intelligence) Anomaly Detector* that identifies abnormalities in dike behavior or sensor malfunctions. The *Reliability Analysis* module calculates the probability of dike failure in case of abnormally high water levels or an upcoming storm and extreme rainfalls. If the failure probability is high then the *Breach Simulator* predicts the dynamics of a possible dike failure, calculates water discharge through the breach and estimates the total time of the flood. After that, the *Flood Simulator* models the inundation dynamics and *Evacuation Simulator* optimizes evacuation routes. Then *Risk Assessment* module calculates flood damage. Finally, *Decision Support System* provides access to different information levels, for experts and citizens. The simulation modules and visualization components are integrated into the Common Information Space [5]. They are accessed from the interactive graphical environment of the multi-touch table or through a web-based application.

* Corresponding author. *E-mail address*: N.Melnikova@uva.nl
* Corresponding author. Tel.: +7-921-319-65-31.

The *Virtual Dike* component runs in parallel with the *Reliability Analysis* module, offering direct numerical simulation to analyze dike stability under specified loadings [9]. The module can be run with a real-time input from water level sensors or with predicted high water levels due to upcoming storm surge or river flood. In the first case, comparison of simulated pore pressures with real data can indicate a change in soil properties or in dike operational conditions (e.g. failure of a drainage pump). In the second case, simulation can predict the structural stability of the dike and indicate the "weak" spots in the dikes that require attention of dike managers and city authorities. Simulated dynamics of dike parameters (including local and overall stability, pore pressure, local stresses and displacements) are compared with sensor data and graphically visualized on monitors of the early warning system. *Virtual Dike* simulation results are then used to feed the AI system with normal and abnormal virtual sensor dynamics [8].

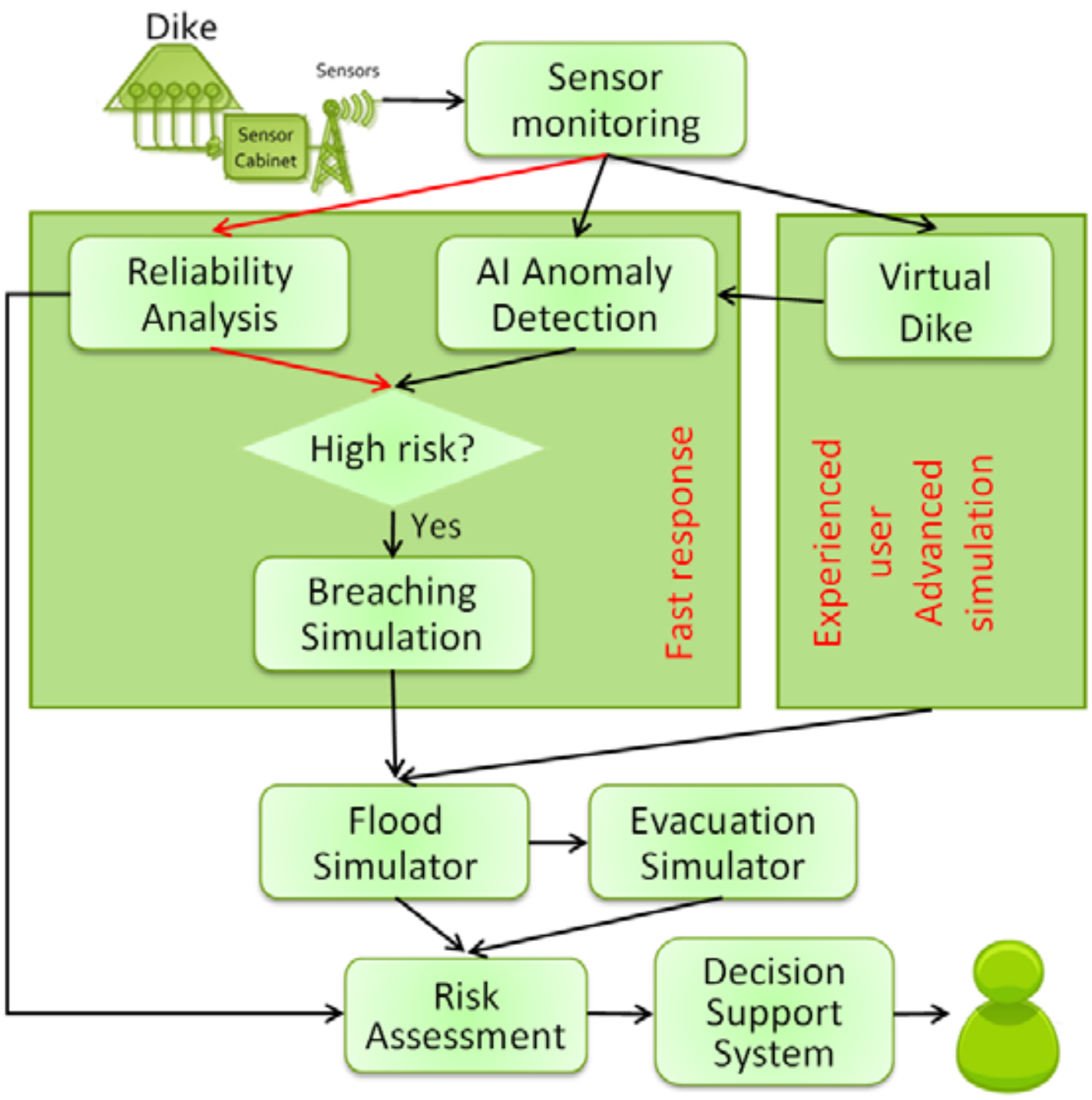

Figure 1. Early warning system workflow

The EWS is described in detail in [5]. In the current paper we only focus on the *Virtual Dike* module design.

Dike stability analyses under hydraulic and structural loads are usually carried either by probabilistic breach analyses based on empirical engineering criteria [13] or by finite element modeling of dike deformation [14]. While the first approach is more robust and is widely used for dike stability analysis, the second approach allows more profound study of physical processes occurring in the dike before the actual failure. Under the frame of the *UrbanFlood* project we create a number of pre-defined and calibrated structural stability analysis models for the dikes connected to the early warning system. Realistic modeling of water flow through the dikes is necessary for correct estimation of effective stresses in the dikes and for predicting their stability. Calibration of diffusivities for the tidal groundwater flow is often performed by tidal methods [15],[16],[17] based on one-dimensional analytical models of semi-infinite or finite aquifers. This method is suitable for aquifers with nearly horizontal phreatic surface. A more accurate way that works well for high amplitude of water level variation is direct numerical simulation. In present work, both analytical and numerical approaches have been tested and compared. Calibration of diffusivities of soil strata has been performed by matching tidal pore pressure fluctuations obtained from numerical simulation and from piezometers installed in several cross-sections of the dike. For heterogeneous soil structures, some averaged and simplified yet heterogeneous soil build-ups have been obtained, so that the response of the dike to the tidal load corresponds well to sensor measurements.
Tidal oscillations of sea level influence the position of phreatic surface in the dike. Moving water table creates the zones with partially saturated soils. Resistance of porous media to the flow is modeled by Darcy's law suitable for low flow velocities [19]. A problem of unconfined porous flow can be solved either by solving Darcy's equation on a moving mesh with adjusting mesh boundary to coincide with surface of zero pore pressure [20], or by using stationary mesh and solving Richards equation with non-linear rheological properties for the media, dependent on the effective water content. These non-linear properties can be modeled by classical models of van Genuchten [22] or Brookes and Corey model [15], as well as by some approximations [21] simplified for faster numerical convergence. We have used Richards' equation with the van Genuchten model, performing simulations on a fixed mesh.

In this paper we present the numerical and analytical results of sensitivity analysis of the porous flow parameters to the variation of soil diffusivity and calibration results performed for the Livedike, an earthen sea dike in Groningen, the Netherlands.

The remainder of this paper is organized as follows: Section 2 contains description of the test case (Livedike) and sensors installed there; Section 3 contains mathematical model description; Section 4 gives model implementation details; Section 5 presents results of porous flow sensitivity analysis; Section 6 contains analytical solutions for one-dimensional saturated porous flow with periodic boundary conditions representing tidal hydraulic load; Section 7 presents the results of diffusivities calibration for the Livedike and Section 8 concludes the paper.

## 2 Livedike: geometry, soil build-up, loadings and sensor data

Livedike is one of the research sites of the UrbanFlood project. It is an earthen sea dike protecting a seaport in Groningen, the Netherlands. The height of the dike is 9 m, the width is about 60 m, the length is about 300 m. The dike has a highly permeable sand core covered by 60 cm clay layer.

(a) (b)

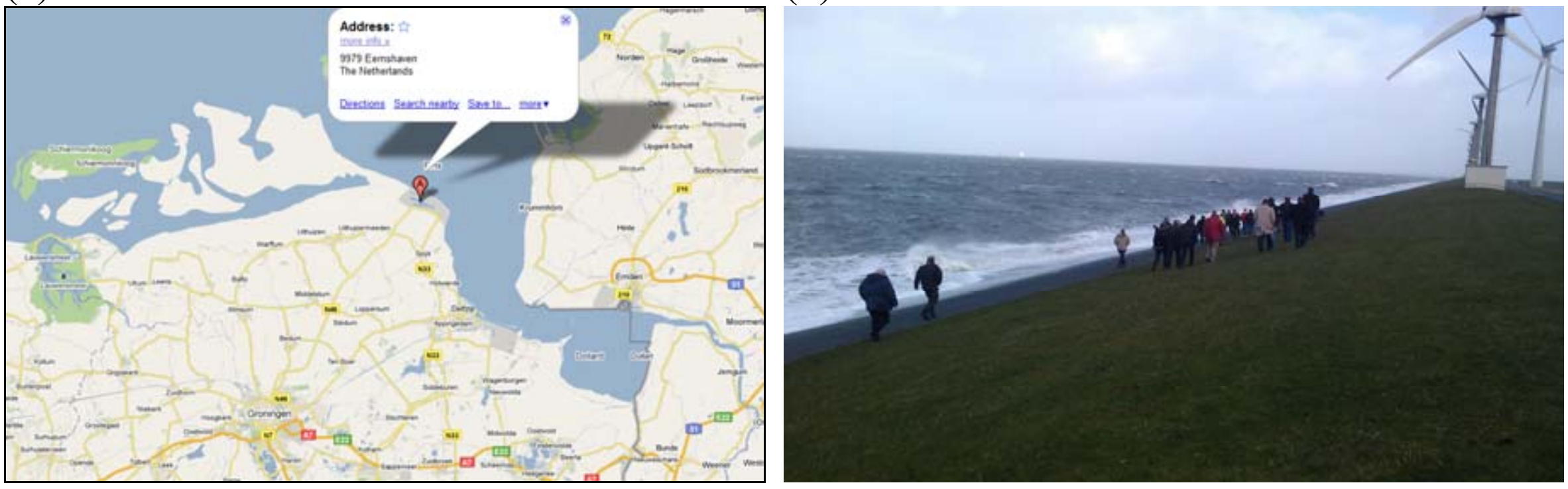

Figure 2. Photo of the LiveDike near Groningen, NL

The Livedike has been equipped with sensors with GPS locations shown in Figure 3(a). Sensors are placed in the four cross-sections, see Figure 3(a,b). These cross-sections have been simulated in 2D models under tidal water loading, in order to calibrate diffusivities, simulate flow through the dike and finally analyze the structural stability of the dike.

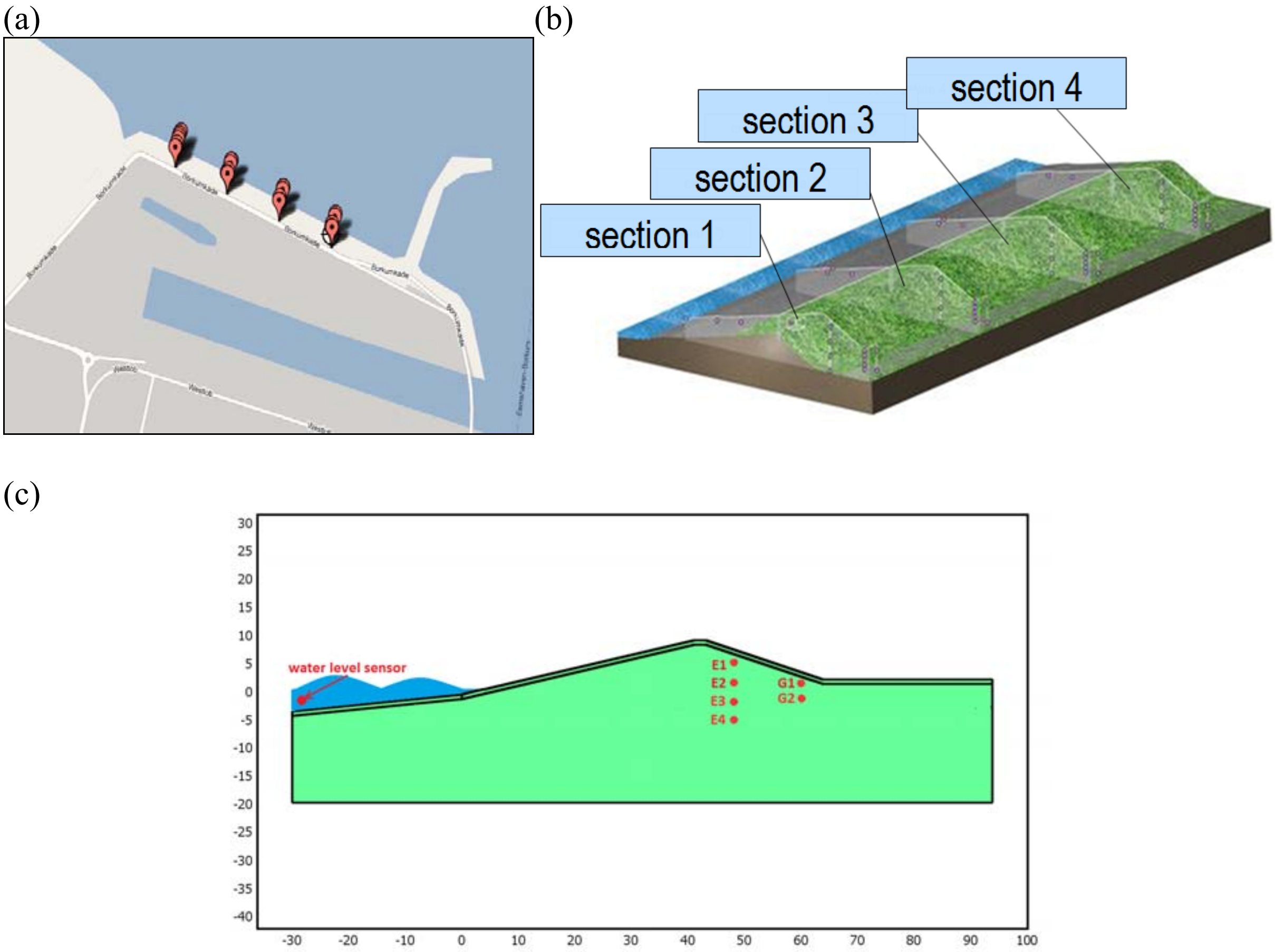

Figure 3. (a) Top view at the Livedike (Eemshaven); (b) LiveDike cross-sections with marked sensor locations; (c) 2D model of dike cross-section with pore pressure sensor locations shown with red dots

A geometric model of dike cross-section with sensor locations is presented in Figure 3(c). Sensors E1-E4 and G1-G2 measure absolute pore pressure and temperature and produce data stream which is available in real-time via a LiveDike Dashboard [7]. For calibration of the model, we have used signals from E3, E4 and G2 pore pressure sensors located below phreatic surface. An input signal for simulation was sea level dynamics coming from water level sensor installed outside of the dike (see Figure 3(c)). Sea-side toe of the dike is located at x=0 m, y=-0.7 m, while national reference sea level is at y=0 m.

The soil build-up for a longitudinal section passing through the crest of the dike is presented in Figure 4. It contains horizontal layers of sand (light orange), silty sand with small clay inclusions (lemon) and 60 cm clay layer that covers the dike (blue). Grey areas are sandy clay. Below the sand layers lies impermeable clay layer (blue). Cone penetration test (CPT) results (cone end resistance and frictional resistance) are schematically shown with black lines. More on CPT testing methodology can be found in [18].

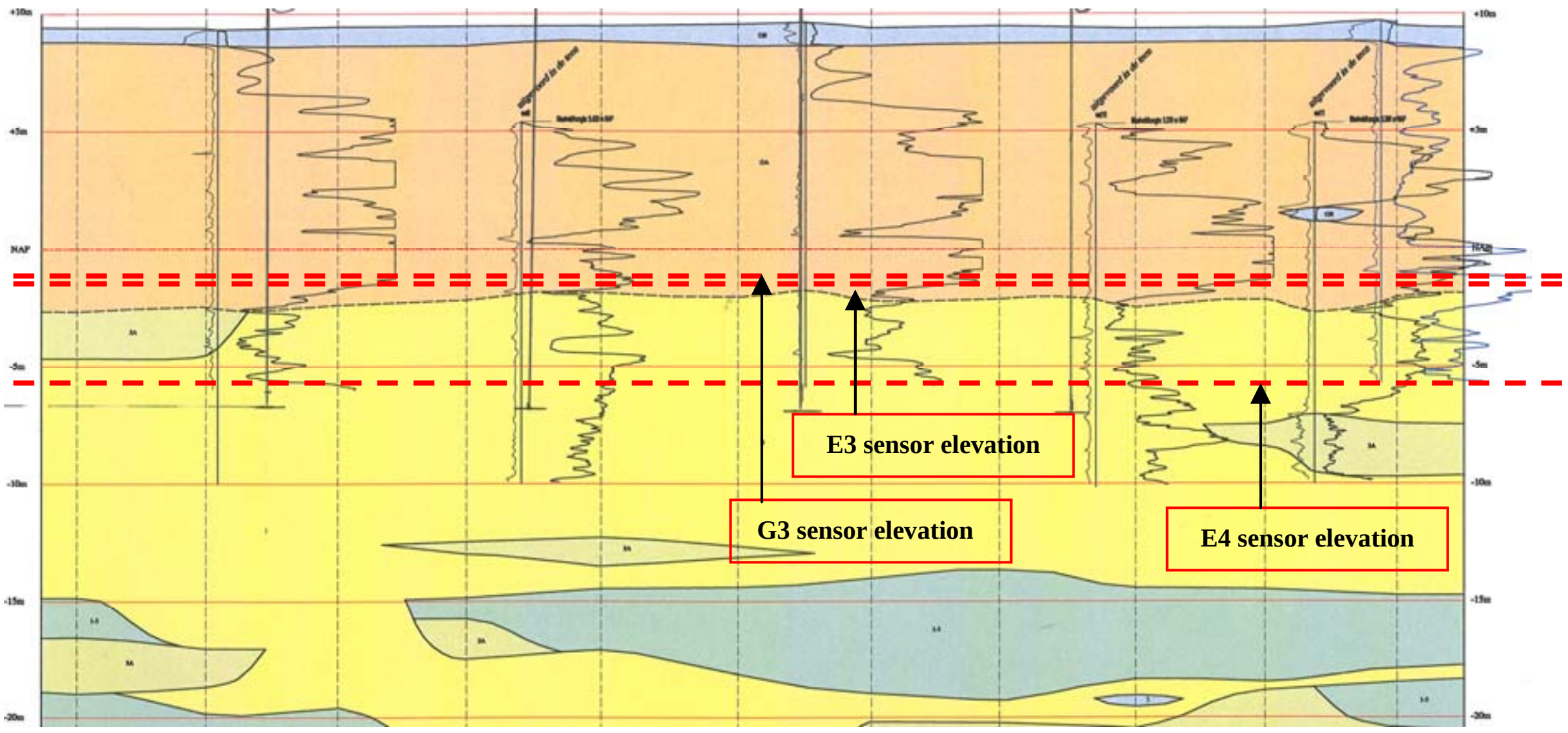

Figure 4. Soil build-up in the Livedike and underlying soil strata

A sample of sensor data showing air pressure, sea level and pore pressure dynamics is presented in Figure 5 and Figure 6, for a time period that has been used for diffusivity calibration ("training period"). Sea level dynamics is presented in Figure 5, with positions of local maximum and minimum marked with dashed lines. Figure 6 presents pore pressure dynamics measured in four sections of the dike. For calibration of diffusivities, the original pore pressure signals have been smoothed by localized linear fit algorithm with adaptive window (the smoothed signals are also shown in Figure 6). Then levels of minimal and maximal tidal pressure have been detected for the smoothed pressure signals. These levels are shown in Figure 6 with horizontal dashed lines, for each sensor. Corresponding pressure values are specified in the legends. Moments of time, corresponding to phases of minimal and maximal pressure values, are marked with vertical dashed lines, for each sensor, with corresponding time values specified in the legends. The obtained relative pressure amplitudes and time delays between local pressure maximum and sea level maximum are presented in Table 1.

Table 1. Livedike pressure sensors measurements: relative pressure amplitudes and time delays between the tide and local pressure fluctuations

| Sea level sensor data | | | | | | |
|---|---|---|---|---|---|---|
| sea level drop<br>258 cm = 253 mbar | | | | time of local maximum<br>9.01.2010 5.00 | | |
| Pore pressure sensors data | | | | | | |
| | Cross - section1 | Cross-section2 | Cross-section3 | Cross - section1 | Cross-section2 | Cross-section3 |
| | Relative daily oscillations amplitude (fraction of tidal daily oscillations amplitude) | | | Time delay between local pressure maximum and sea level maximum, minutes | | |
| E4 | 0.21 | 0.11 | 0.10 | 18 | 3 | 18 |
| E3 | 0.09 | 0.10 | 0.07 | 24 | 9 | 38 |
| G2 | 0.03 | 0.08 | 0.04 | 49 | 19 | 9 |

E3 and E4 sensors are located at the same distance from sea (x=50 m), but at different levels (y=-1.5 m and y=-5.5 m from reference level, correspondingly). E3 pressure oscillations are lower than E4 oscillations and this fact points to the presence of vertical heterogeneity in the dike. Time delay between E4 oscillations (at x=50 m) and tidal oscillations (at x=0 m) varies in the range between 3 and 18 min, which indicates highly permeable sand in the zone 0 <x<50 m. E3 oscillations lag from tidal oscillations by 9-38 min.

(a)

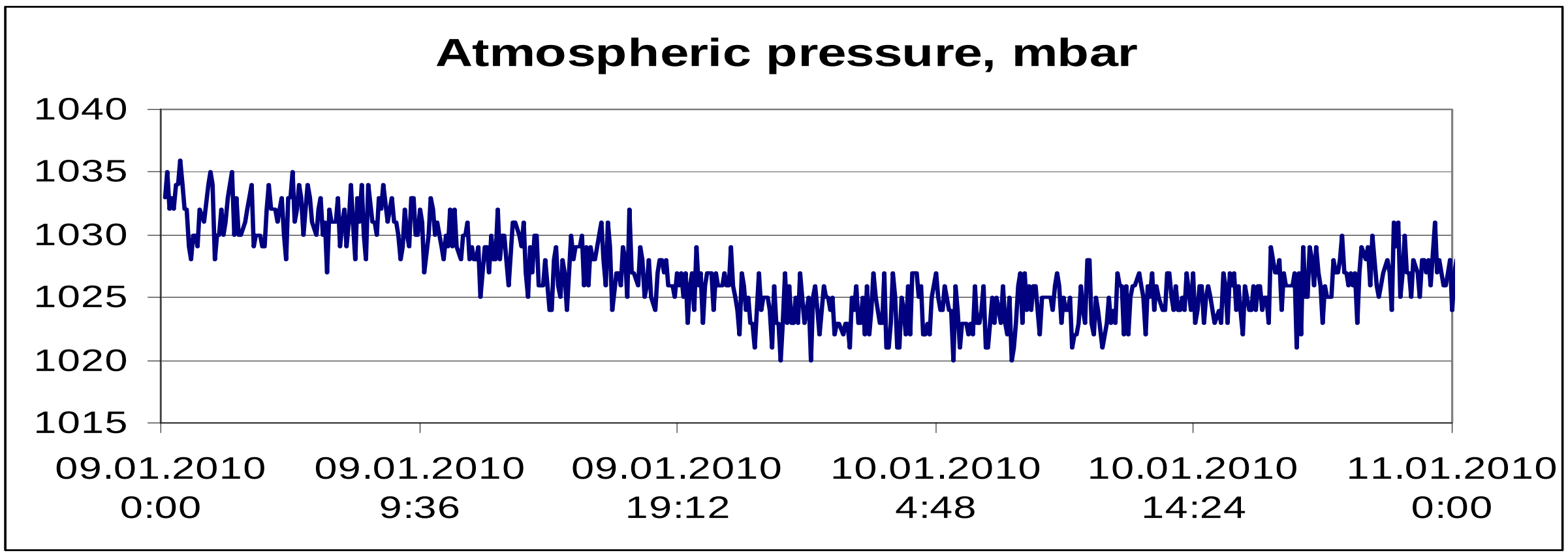

(b)

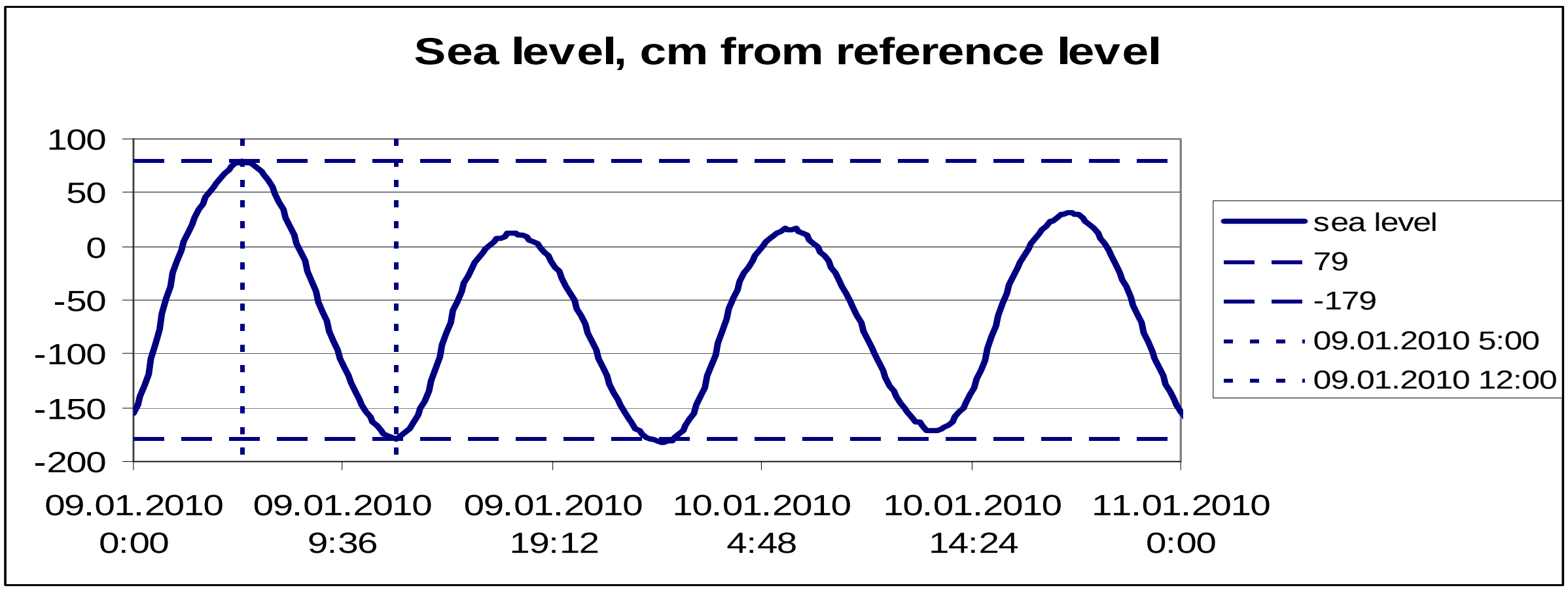

Figure 5. Livedike: (a) atmospheric pressure [mbar] and (b) sea level [cm] registered by sensors.

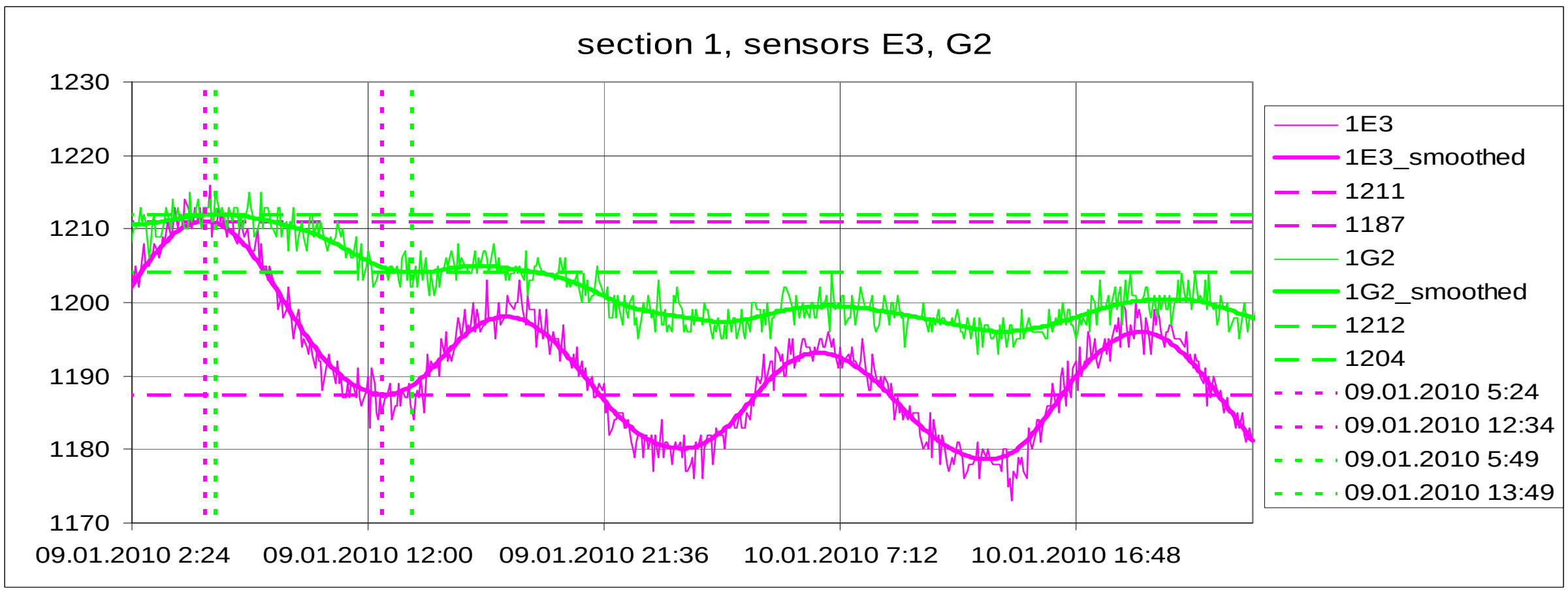

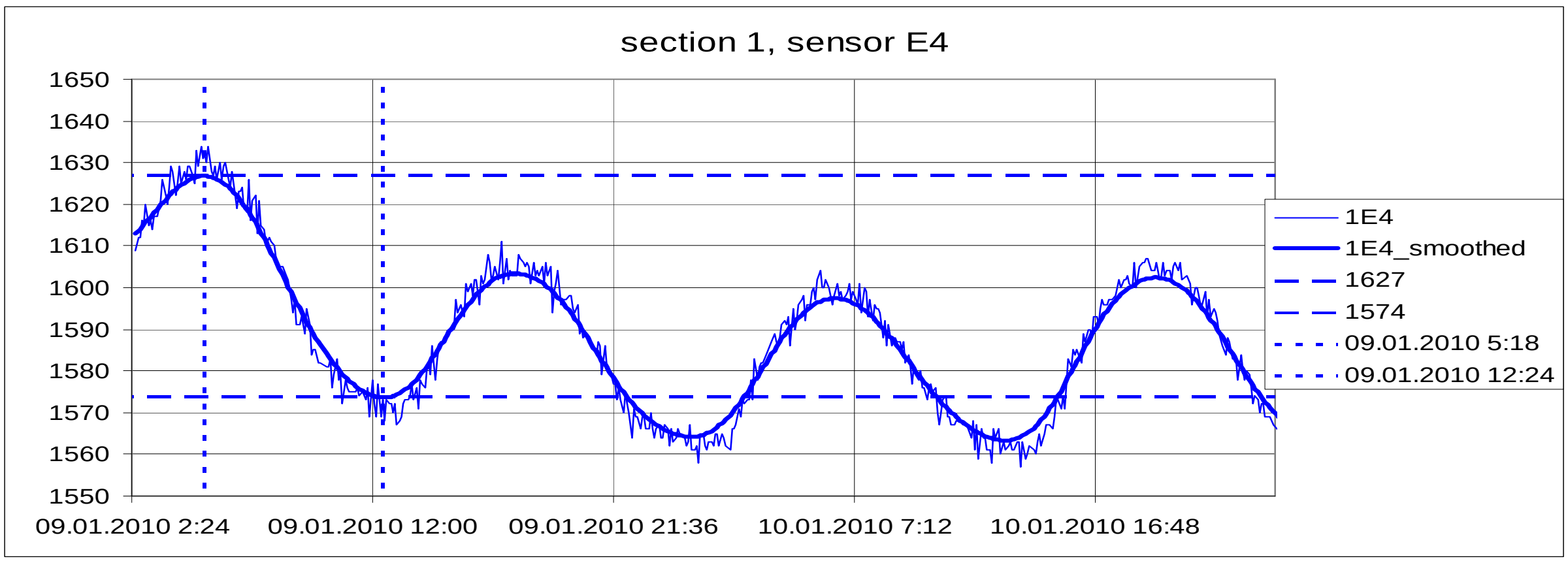

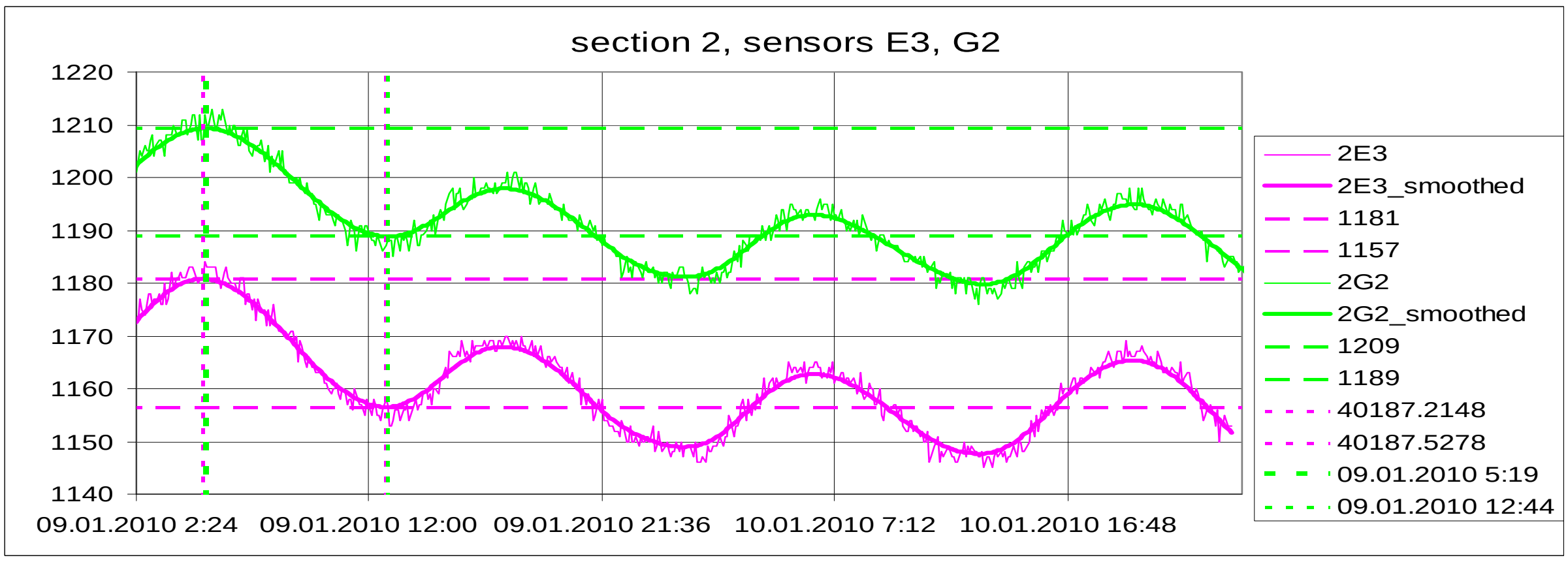

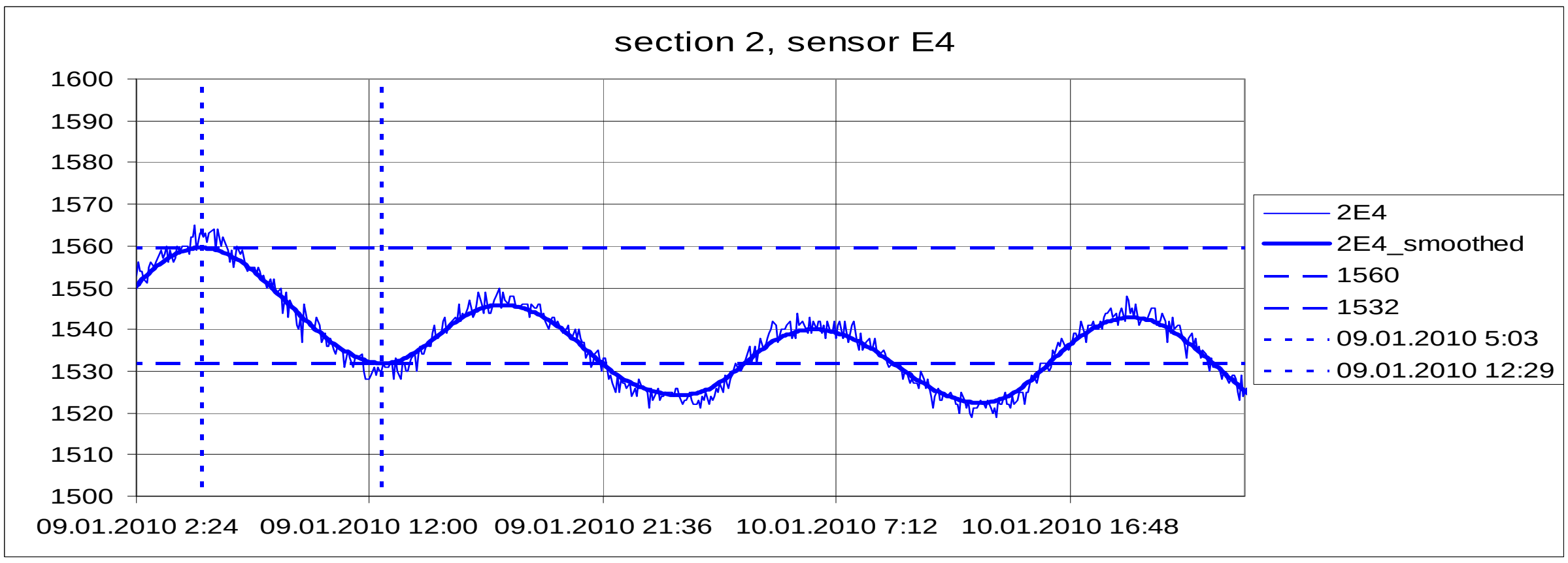

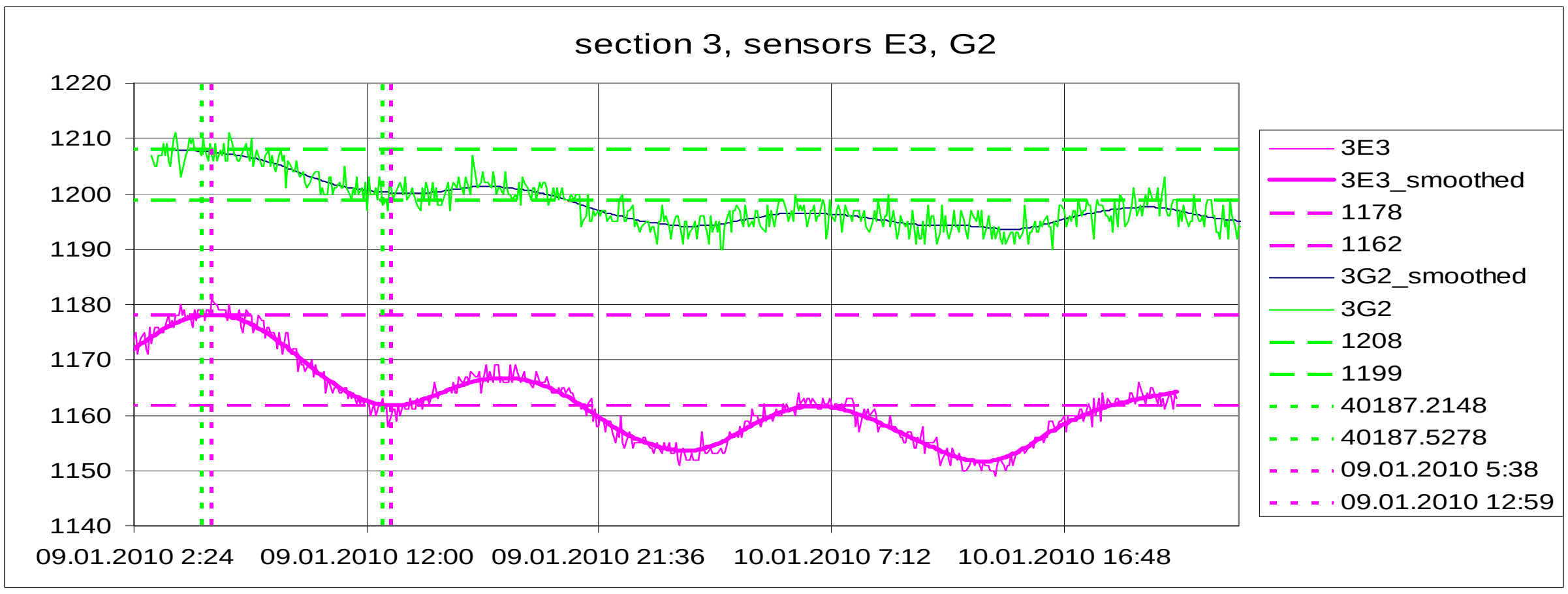

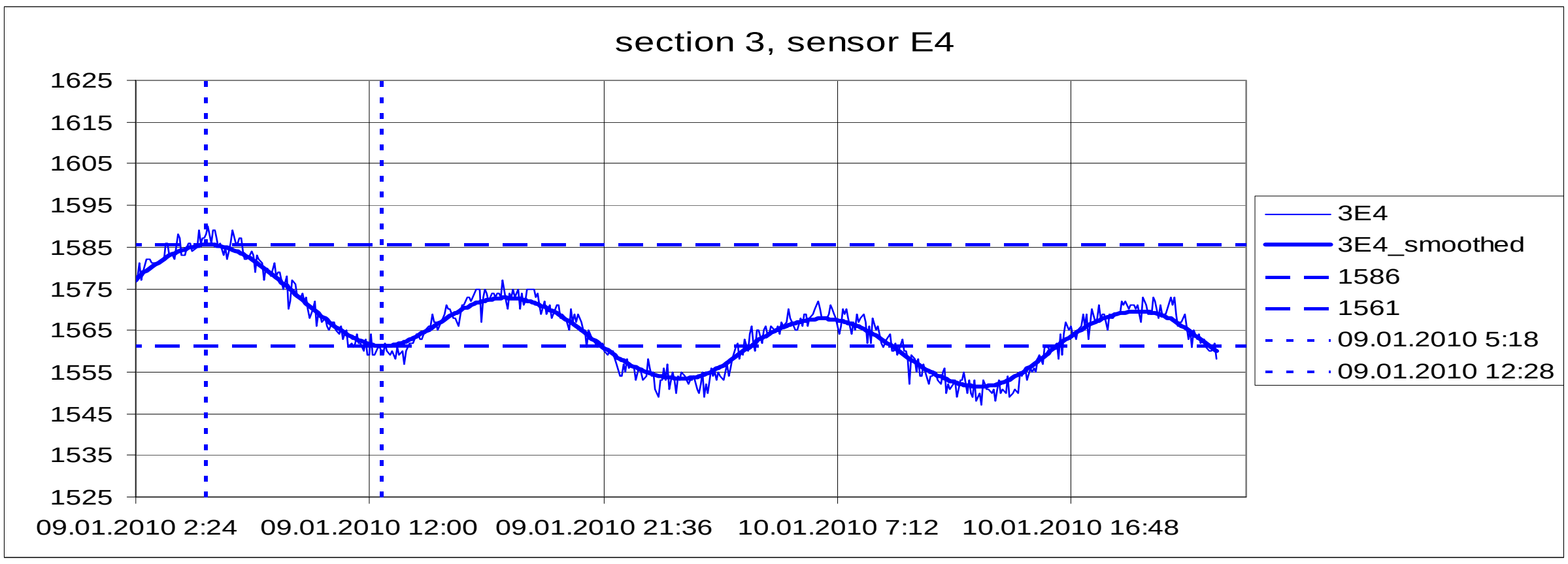

Figure 6. Livedike: absolute pore pressure [mbar] registered by sensors and smoothed pressure signals

E3 and G2 sensors are located at approximately the same level (-1.2 m÷-1.5 m from ref. level), but at different distances from the sea (x=50 and x=62 m, correspondingly).The amplitude of pore pressure dissipates quickly within 12 m of horizontal distance between E3 and G2 sensors, and that fact indicates presence of horizontal heterogeneity in sand layers, with diffusivity decreasing with the distance from the sea, up to a dense impermeable zone near G2. It is observed in all four sections, and at most in the first section (Table 1). This impermeable zone creates high time lag between G2 oscillations and tidal oscillations: the lag varies in the range from 19 to 49 min for different cross-sections due to significant dike heterogeneity.

We have built a heterogeneous 2D model of the 1$^{st}$ cross-section of the Livedike, in order to reproduce actual pore pressure dynamics. In the following sections we present the mathematical model of porous flow, numerical and analytical study of diffusivity influence on the pore pressure dynamics in the dike and, finally, calibration of diffusivities for the Livedike.

## 3 Mathematical model

Water flow through the dike is described by Richards' equation with the van Genuchten model for water retention in partially saturated soil around the phreatic surface (1).

$$(C+\theta_e S)\frac{\partial p}{\partial t}+\nabla\cdot[-\frac{K_S}{\mu}k_r\nabla(p+\rho g z)]=0 \qquad (1)$$

where $C$, $\theta_e$, $S$ denote specific moisture capacity [1/Pa], effective water content and specific storage [1/Pa], respectively; $S=1/K$, where $K$ is soil skeleton bulk modulus; $p$ is water pressure (negative in unsaturated zone), [Pa]; $t$ is time [s]; $K_S$ is permeability of saturated media [m$^2$]; $k_r=k_r(p)$ is relative permeability; $\mu$ is dynamic viscosity of water [Pa·s], ($\mu$ is a function of water temperature and changes its value during the year); $g$, $\rho$, $z$ are standard gravity [m/s$^2$], water density [kg/m$^3$] and vertical elevation coordinate [m], respectively. Specific moisture capacity $C$ and relative permeability $k_r$ are described by van Genuchten equations [22]:

$$C=\frac{\partial\theta}{\partial p}=\begin{cases}\frac{am}{1-m}(\theta_s-\theta_r)\theta_e^{1/m}\left(1-\theta_e^{1/m}\right)^m, & p<0,\\ 0, & p\ge 0\end{cases}$$
$$k_r=\begin{cases}\theta_e^{l}\left[1-(1-\theta_e^{1/m})^m\right]^2, & p<0\\ 1, & p\ge 0\end{cases} \qquad (2)$$

where $\theta$ is water content; $\theta_s$ and $\theta_r$ are saturated and residual water content, specific for each soil.

Effective water content is calculated as

$$\theta_e = \begin{cases} \dfrac{1}{(1+(a|p/\rho g|)^n)^m}, & p<0 \\ 1, & p \ge 0 \end{cases}, \qquad (3)$$

where *a, n, m=1-1/n, l* are van Genuchten parameters specific for each soil type (see Table 2 for these parameters values).

For the Livedike, a planar cross-section has been modeled, with the boundary conditions specified as follows (Figure 7):

magenta boundaries are walls with zero normal flux $\dfrac{\partial p}{\partial n} = 0$;

black boundaries are sea side with tidal pressure oscillations specified:

$$\begin{cases} p = \rho g \cdot (h(t) - y) \;\; for \;\; y \le h(t) \\ p = 0 \;\; for \;\; y > h(t) \end{cases}, \qquad (4)$$

where *h(t)* is oscillating sea level [m], measured by sensors or predicted by hydrological model;

blue boundaries are land side with attenuated oscillations of ground water level:

$$\begin{cases} p = \rho g \cdot (h_{gw}(t) - y) \;\; for \;\; y \le h_{gw}(t), \\ p = 0 \;\; for \;\; y > s(t) \end{cases}, \qquad (5)$$

where $h_{gw}(t)$ denotes oscillating ground water level, representing attenuated and altered tidal signal.

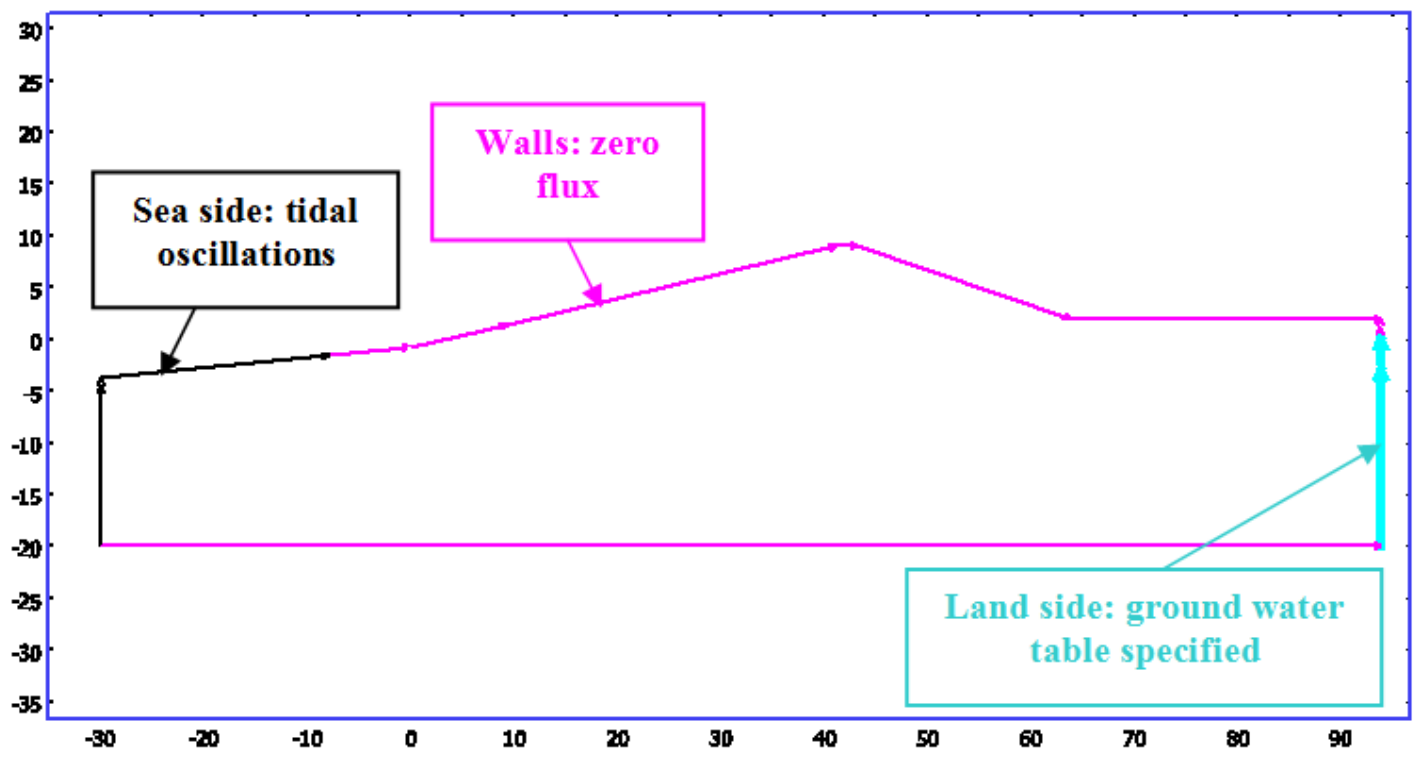

Figure 7. 2D simulation domain and boundary conditions

In the regime of forced tidal oscillations the initial condition in (1) does not affect the steady solution, due to dissipation of initial pore pressure distribution within several tidal periods. We have specified hydrostatic distribution below *y*=0 m as initial condition:

$$\begin{cases} p = -\rho g y \;\; if \;\; y \le 0, \\ p = 0 \;\; if \;\; y > 0 \end{cases}$$

In saturated zone, where $\theta_e=1$, $C=0$, $k_r=1$, porous flow is modeled by linear and parabolic Laplace equation:

$$\frac{\partial p}{\partial t}+\nabla\cdot[-\frac{K_S}{S\mu}\nabla(p+\rho gy)]=0, \tag{6}$$

where $d=\frac{K_S}{S\mu}$ [$m^2$/s] is soil-water diffusivity, the only soil parameter that influences pore pressure dynamics under the specified load.

The structural sub-model describes deformation dynamics of the dike under hydraulic load, gravity and volumetric pore pressure load obtained from flow simulation. Linear elastic – perfect plastic deformations of soil skeleton are described by general equations of plastic flow theory [23]:

$$\begin{cases} \nabla\cdot\underline{\underline{\sigma}}-\nabla p+\rho_S \underline{g}=0 \\ \underline{\underline{\sigma}}=\frac{E}{1+\nu}\left[\frac{\nu}{(1-2\nu)}\varepsilon\underline{\underline{I}}+\underline{\underline{\varepsilon}}\right], \ if \ \ F<0, \\ \underline{\underline{\dot{\varepsilon}}}_{pl}=-q\frac{\partial F}{d\underline{\underline{\sigma}}}, \ \ \underline{\underline{\dot{\varepsilon}}}=\underline{\underline{\dot{\varepsilon}}}_{pl}+\frac{1}{3K}I_1\underline{\underline{I}}, \ if \ \ F=0 \end{cases}, \tag{7}$$

where $\nabla=\underline{e}_x\frac{\partial}{\partial x}+\underline{e}_y\frac{\partial}{\partial y}+\underline{e}_z\frac{\partial}{\partial z}$ is the Hamiltonian; $\rho_s$ is soil density; $\underline{g}$ is gravity vector; $\underline{\underline{\sigma}}$ is effective stress tensor (compressive stresses are negative); $E$ is Young's modulus; ν is Poisson's ratio; $\varepsilon=\varepsilon_{xx}+\varepsilon_{yy}+\varepsilon_{zz}$ is volume deformation (positive for expansion); $\underline{\underline{I}}$ is unit tensor; $\underline{\underline{\varepsilon}}=(\nabla\underline{U}+(\nabla\underline{U}^T)/2$ is deformation tensor; $\underline{U}$ is vector of displacements; $\underline{\underline{\dot{\varepsilon}}}_{pl}$ is plastic deformation rate tensor; q is plastic multiplier; $F$ is plastic yield function; $K=\frac{E}{3(1-2\nu)}$ is bulk modulus; $I_1=\sigma_{xx}+\sigma_{yy}+\sigma_{zz}$ is the first stress invariant.

Plastic flow in structure has been modeled with a modification of Drucker-Prager plasticity model, optimized for plane strain problems by providing the best approximation of Mohr-Coulomb surface in stress space for 2D cases [27]:

$$F=\alpha\cdot I_1+\sqrt{J_2}-F_{DP},$$

where $J_2=I_1^2/3-I_2$ is second deviatoric stress invariant, $I_2=\sigma_{xx}\cdot\sigma_{yy}+\sigma_{zz}\cdot\sigma_{yy}+\sigma_{xx}\cdot\sigma_{zz}-\sigma_{xy}^2$ is second stress invariant; $\alpha$ and $F_{DP}$ are constants: $\alpha=tg(\varphi)/\sqrt{9+12\cdot tg^2(\varphi)}$, $F_{DP}=3c/\sqrt{9+12\cdot tg^2(\varphi)}$; $c,\varphi$ are cohesion and internal friction angle, respectively.

Boundary conditions for structural sub-model of the Livedike are specified as follows:

- roller condi\\tion at the vertical borders: $U_x=0$, $\tau_{xy}=0$;
- fixed base of the dike: $U_x=U_y=0$ at the bottom horizontal border;
- normal pressure acting below transient sea level at the slopes of the dike:

$$\begin{cases} n\cdot\underline{\underline{\sigma}}=-\rho g\cdot(h(t)-y) \ \ if \ \ y\le h(t), \\ n\cdot\underline{\underline{\sigma}}=0 \ \ if \ \ y>waterlevel(t) \end{cases}$$

Differentiation in (7) is performed with respect to pseudo-time with initial condition $\underline{\underline{\varepsilon}}_{pl}=0$.

Equations (1) and (7) describe one-way coupled fluid-structure interaction system. In (1) we did not take into account squeezing/suction of pore water with the volume deformation of pores (source intensity is zero in (1)). This assumption has been made for the Livedike as it constructed from sand which usually develops minor pore compaction/expansion.

Table 2 gives a list of soil parameters that have been set for the Livedike using reference properties of sand.

Table 2. Livedike – soil parameters, including calibrated diffusivities

| Van Genuchten parameters | | | Young's modulus $E$, Pa | Poisson's ratio $\nu$ | Friction angle $\varphi$, degrees | Cohesion $c$, Pa |
|---|---|---|---|---|---|---|
| $\alpha$, [1/m] | $n$ | $l$ | | | | |
| 8 | 1.5 | 0.5 | $10^{10}$ | 0.3 | 30 | 0 |

Water viscosity has been defined (for simplicity) as a step function of water temperature (Table 3). Dynamics of water temperature during a year cycle are presented in Figure 8, for year 2009. Due to variation of water viscosity, value of soil diffusivity in July is 1.8 times higher than in January.

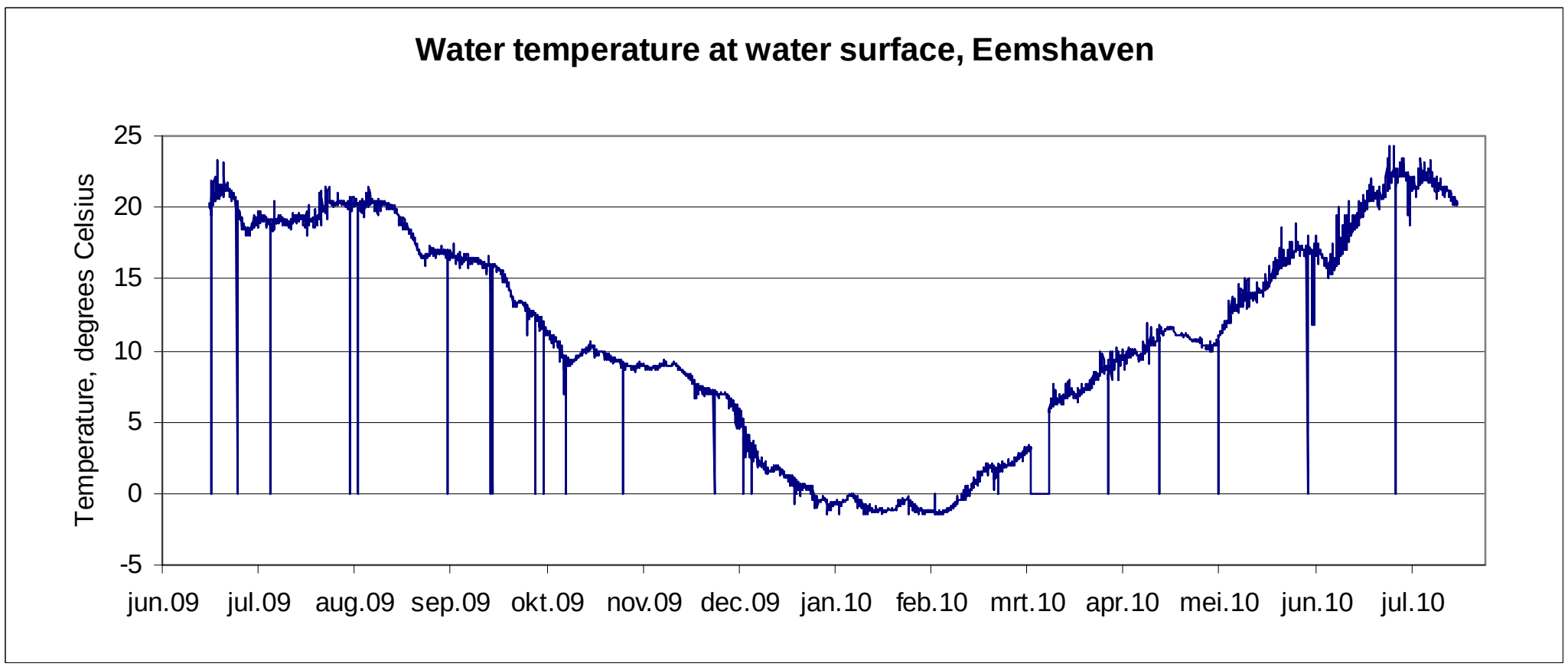

Figure 8. Sea water temperature distribution during year cycle

Table 3. Water viscosity values

| Temperature, °C | Dynamic viscosity, Pa·s |
|---|---|
| 20 | $1.004 \cdot 10^{-3}$ |
| 10 | $1.307 \cdot 10^{-3}$ |
| 0 | $1.797 \cdot 10^{-3}$ |

## 4 Implementation details and model parameters

In the *Virtual Dike* module, equations (1) and (7) are solved using the finite element package COMSOL 3.5a on a finite element mesh composed of triangle elements with second order of space approximation. Time integration is performed by implicit backward second order method. Newton-Raphson iteration scheme is used for solving nonlinear algebraic equations at each time integration step. During Newton-Raphson iterations, systems of linear algebraic equations are solved by direct PARDISO solver from BLAS library.

The *Virtual Dike* module has been integrated into the Common Information Space (CIS) [5] on the platform of national Dutch supercomputing system SARA [28]. The module runs in real time mode, receiving water level sensor signal as input data and producing "virtual sensors" dynamics (flow and structure parameters).

To start simulation, CIS launches Linux Ubuntu virtual machine with *Virtual Dike* on it on SARA and writes sensor input (sea level value) to the specified directory, in real-time (each 5-10 minutes). The output from *Virtual Dike* is stored in a specified directory on a hard drive, from where it is accessed by CIS, compared to sensor measurements and visualized at the front-end.
Within *Virtual Dike* module, automatic sensor input is implemented by running MATLAB script, which monitors the input directory for new input files, reads input data, starts COMSOL simulation and stores virtual sensor output (see Figure 9 for the internal *Virtual Dike* simulation workflow).

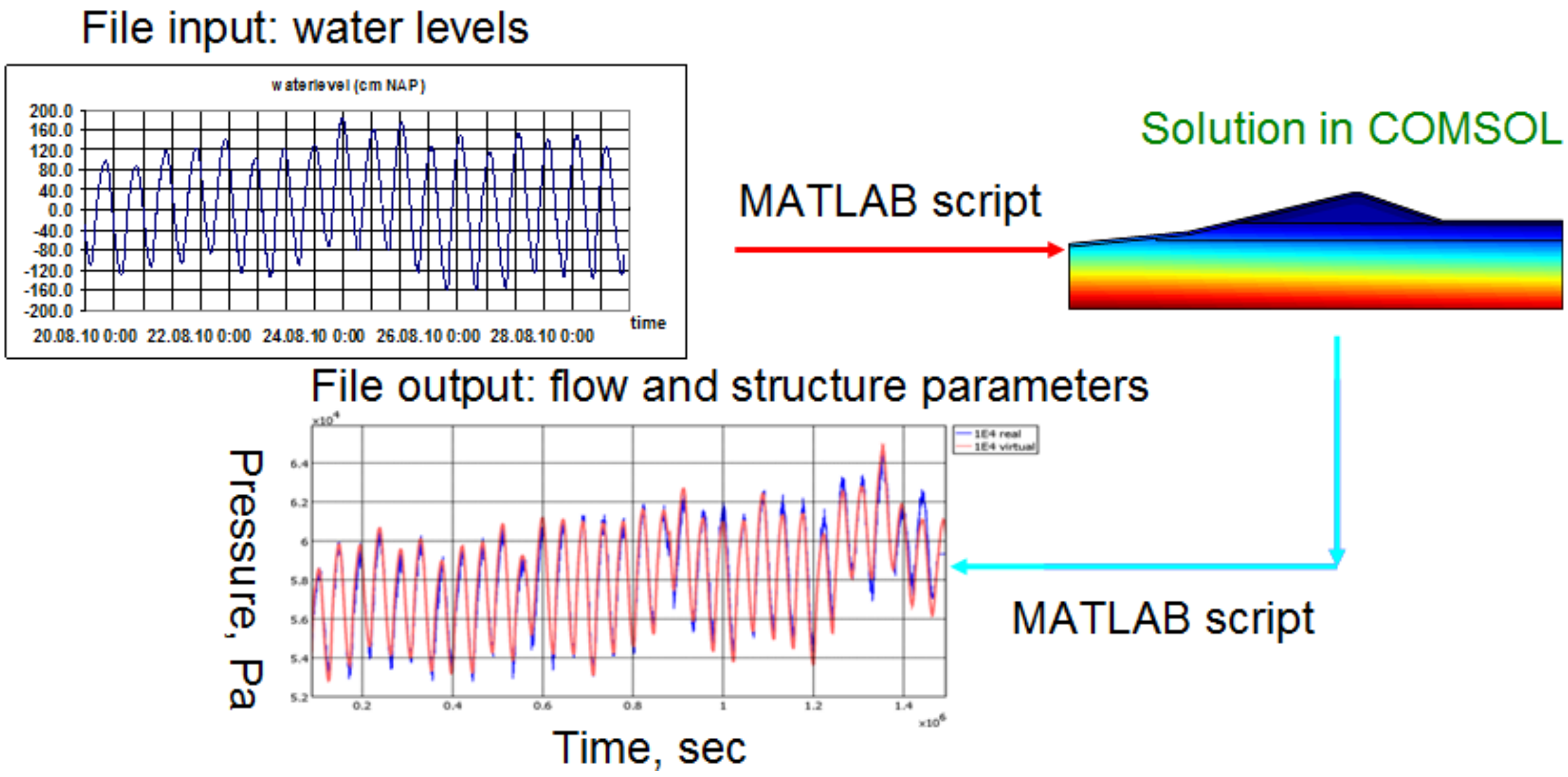

Figure 9. Virtual dike simulation workflow

## 5 Sensitivity analysis of pressure amplitude and time delay to the variation of soil diffusivity

Sensitivity analysis has been performed to study the influence of saturated soil-water diffusivity on pore pressure dynamics in the dike. 2D homogeneous dike model has been considered. Geometric prototype of the model is Livedike's cross-section. Boundary conditions zones have been described in section 3. At the seaside harmonic tidal pressure oscillations are specified; at the landside constant ground water level is specified (zero meters from average sea level). A number of porous flow simulations have been performed, with saturated diffusivities varied in the range of 0.1-1000 $m^2/s$. Water viscosity was constant: $\mu$=0.001 Pa·s. Distribution of relative pore pressure amplitudes, normalized to tidal amplitude, is presented in Figure 10(a), for a horizontal slice of the dike (at the level $y = -5.5$ m).
For relatively high values of diffusivity ($d = 10 \div 1000\, m^2/s$) relative pressure amplitude distribution is linear with a very small non-linear tail close to the sea-side (left) slope. The non-linear part corresponds to the zone where the flow is essentially two-dimensional: at $x \leq 0$ m, water penetrates into the domain both through the vertical boundary and through the under-water slope of the dike (see Figure 11 for arrow plot of flow velocity). At $x \geq 0$ m the flow direction behaves almost like one-dimensional, and relative pressure amplitude distribution corresponds to the 1D analytical solution presented below (sec. 6).

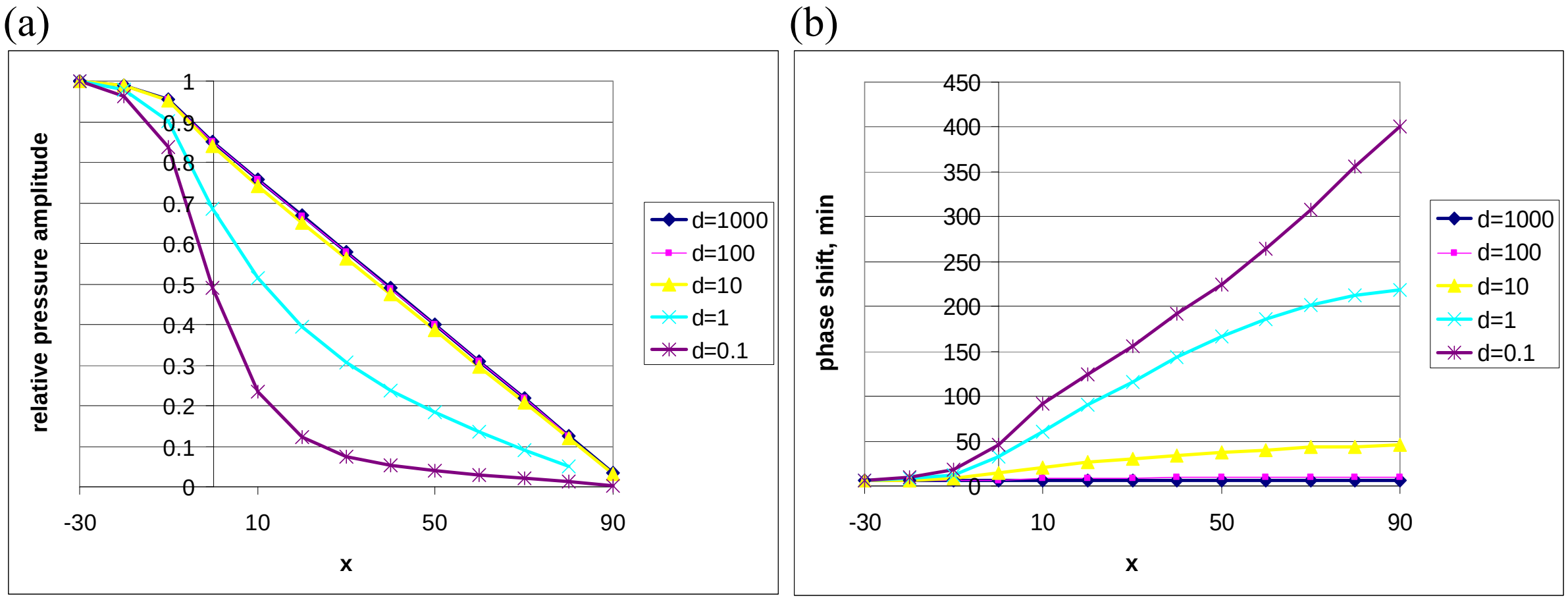

Figure 10. (a) Relative pressure amplitude distribution along the dike. (b) Time delay distribution along the dike. Data shown in a horizontal slice $y$ = - 5.5 m (at the level of the E4 sensor, see Figure 3(b))

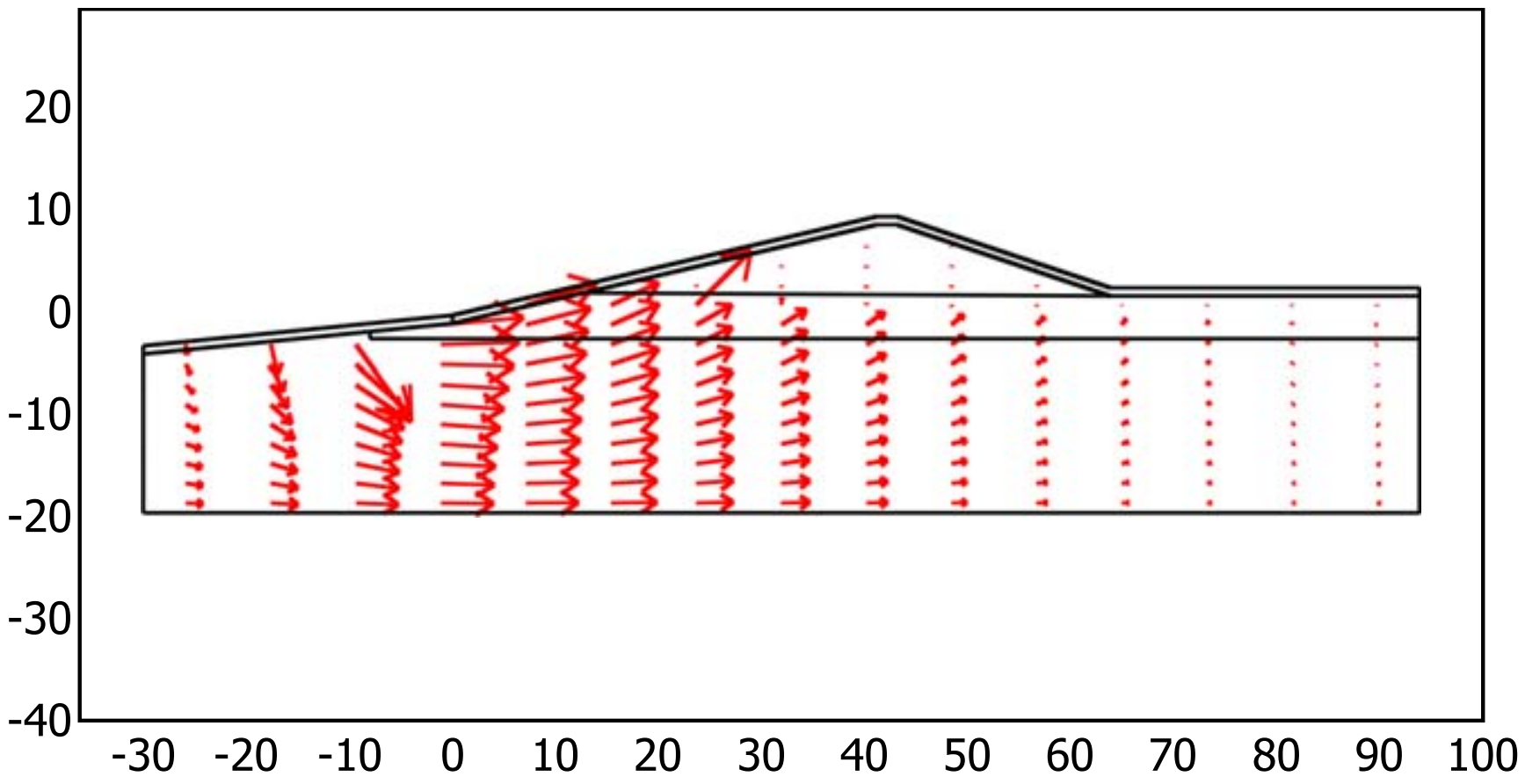

Figure 11. Arrow plot of velocity field (high tide).
The flow is essentially two-dimensional at $x \leq 0$ m

Figure 10(a) clearly shows that relative pressure amplitude for loose, permeable media (like gravel and coarse sand) is insensitive to the actual value of diffusivity (lines for d=1000, d=100, d=10 coincide). This linear distribution is only defined by the amplitude of sea oscillations and by length of the domain. To the contrary, the time delay is sensitive to the value of diffusivity in the whole range (Figure 10(b)), therefore time delay can be calibrated by tuning diffusivity value.

For diffusivities $d \leq 1$ m$^2$/s, significant non-linearity appears in pressure amplitude distribution: pressure amplitude within the dike depends on the diffusivity.

Figure 12 presents pressure amplitude and time delay as functions of diffusivity, in Livedike E4 sensor location (50;-5.5). The amplitude and phase delay values of E4 sensor are shown in Figure 12 with dashed lines. From Figure 12 it is clear that matching both the amplitude and phase lag with only one parameter (diffusivity) is impossible: matching the amplitude value requires diffusivity $d \sim 1$ m$^2$/s, while matching the time delay requires that $d \sim 100$ m$^2$/s. Formally, besides diffusivity, one more parameter is necessary to match data for one sensor. In fact, this contradiction indicates presence of heterogeneity in the Livedike soil build-up (while the prototype dike in sensitivity analysis is homogeneous). Thus we construct a model of a dike as a set of horizontal slices, each slice divided into a number of homogeneous sectors with constant diffusivity. The length of a sector is the second necessary parameter for matching sensor data (Figure 13). Figure 13 presents a scheme of construction of heterogeneous dike model to match sensor data. Sensors $E_1$, $E_2$, $G_1$ are not taken into consideration in the model as they are located above phreatic surface and they do not produce data on pore pressure. For 6 values to match (these are pressure amplitude and time delay for 3 sensors: $E_4$, $E_3$, $G_2$), 6 parameters have been used: lengths of homogeneous zones $L_1$, $L_2$ and diffusivities $d_1$, $d_2$, $d_3$, $d_4$ (see Figure 13). After calibration total length of simulation domain equals to the sum of parameters $L_1$, $L_2$.

(a) (b)

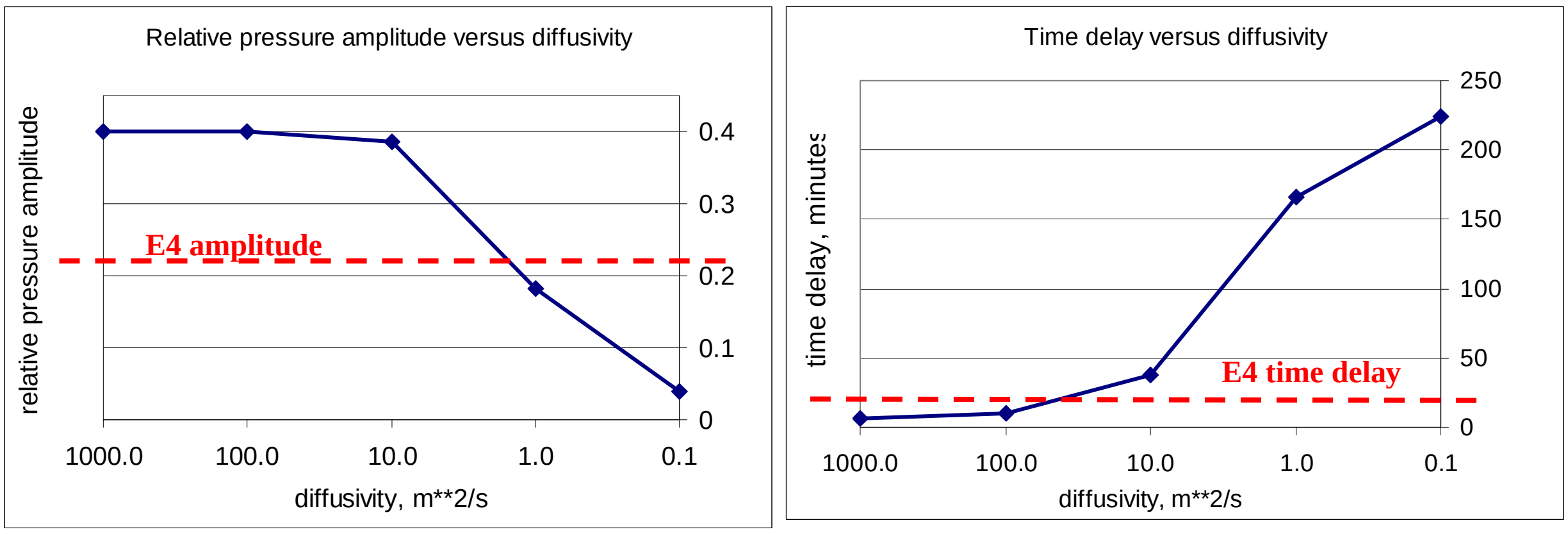

Figure 12. Influence of soil diffusivity on (a) relative pore pressure amplitude and (b) time delay in E4 sensor location (50m; -5.5m). Actual values shown with dashed lines

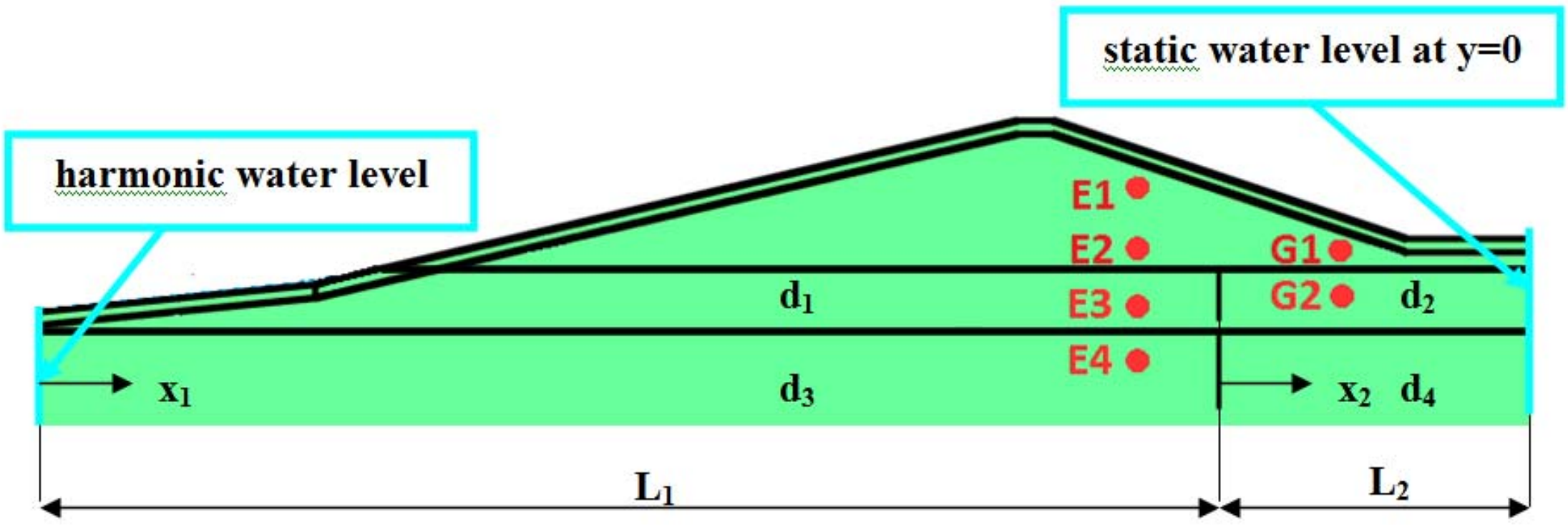

Figure 13. Construction of a heterogeneous dike model to match sensor data

## 6 Analytical analysis of tidal propagation in one-dimensional homogeneous aquifer

In this section two analytical solutions for the problem of harmonic flow in one-dimensional saturated homogeneous aquifer are derived and compared to direct numerical solutions which were discussed in section 5. One-dimensional analytical model can be used for modeling tidal propagations through the aquifers with low gradient of phreatic line.

The objectives for employing one-dimensional analytical models for dike diffusivity calibration are:

- obtaining formulas for initial guess values of diffusivity;
- qualitative study of penetration of tidal waves through the dike.

Flow in one-dimensional saturated aquifer is described by the equation

$$\frac{\partial p}{\partial t} - d \cdot \frac{\partial^2 p}{\partial x^2} = 0\,, \qquad (8)$$

Harmonic boundary conditions defining two problems are considered:

- Semi-infinite aquifer with sine oscillations of water pressure at the boundary x=0:

$$p(x,t)\big|_{x=0} = A\sin(\omega t), \quad p(x,t)\big|_{x\to\infty} \to 0 \tag{9}$$

where $A$ is amplitude of pressure oscillations; $\omega$ is angular frequency;

- Finite aquifer with sine pressure oscillations at $x=L$ and constant pressure $p=0$ at $x=0$:

$$p(x,t)\big|_{x=L} = A\sin(\omega t), \quad p(x,t)\big|_{x=0} = 0 \tag{10}$$

The solution for the semi-infinite aquifer problem (8) with boundary conditions (9) is expressed as follows [25]:

$$p(x,t) = Ae^{-x\sqrt{\frac{\omega}{2d}}}\cdot\sin\left[\omega(t - x\sqrt{\frac{1}{2d\omega}})\right], \tag{11}$$

It represents a wave of pore pressure travelling in compressible soil, with the amplitude $p_A$ [Pa] dissipating exponentially with the distance from the inlet, and time delay $\Delta t$ [s] growing linearly with the distance:

$$p_A(x) = Ae^{-x\sqrt{\frac{\omega}{2d}}}, \quad \Delta t = x\sqrt{\frac{1}{2d\omega}} \tag{12}$$

Applying solution (11) to the model of the dike described in section 5 (for $-30 \le x \le 90$) we get distributions of relative pressure amplitude $p_A(x)/A$ and time delay $\Delta t$ (in logarithmic scale), presented in Figure 14(a,b). Diffusivity $d$ varied in the range between 0.1 and 1000 $m^2/s$. Tidal frequency $\omega=2\pi/T$, where $T$=12 hrs 25 min.

Figure 14(a) gives an estimate for a distance of tidal waves penetration in homogeneous aquifer. For dense impermeable soils with diffusivity d≤0.1 pressure amplitude dissipates to a level of 4% of tidal amplitude, within the distance of 120 meters from the sea. For highly permeable soils with diffusivity ≥10$m^2/s$, pressure amplitude distribution is linear in the whole domain, and this linear distribution has been confirmed by 2D numerical analysis (section 5).
According to formula (12), slow seasonal water table fluctuations propagate further into an aquifer than daily fluctuations do, and this was taken into consideration for Livedike when specifying land side boundary conditions in porous flow problem (section 7).

For the finite aquifer problem (8) with boundary conditions (10), solution representing steady harmonic oscillations and satisfying zero boundary condition $p(x,t)\big|_{x=0} = 0$ can be expressed as sum of two complex conjugated independent partial solutions of (8):

$$p = Ce^{i\omega t}\sinh(\sqrt{\frac{i\omega}{d}}x) + \overline{C}e^{-i\omega t}\sinh(\sqrt{-\frac{i\omega}{d}}x), \tag{13}$$

where $i = \sqrt{-1}$; $C = \mathrm{Re}(C) + i\cdot\mathrm{Im}(C)$ is a complex constant to be determined from the harmonic boundary condition:

$$\begin{aligned} &p(x,t)\big|_{x=L} = A\sin(\omega t) \Leftrightarrow \\ &\Leftrightarrow C(\cos(\omega t) + i\cdot\sin(\omega t))\sinh(\sqrt{i\frac{\omega}{d}}L) + \overline{C}(\cos(\omega t) - i\cdot\sin(\omega t))\sinh(\sqrt{-i\frac{\omega}{d}}L) = A\sin(\omega t) \end{aligned}, \tag{14}$$

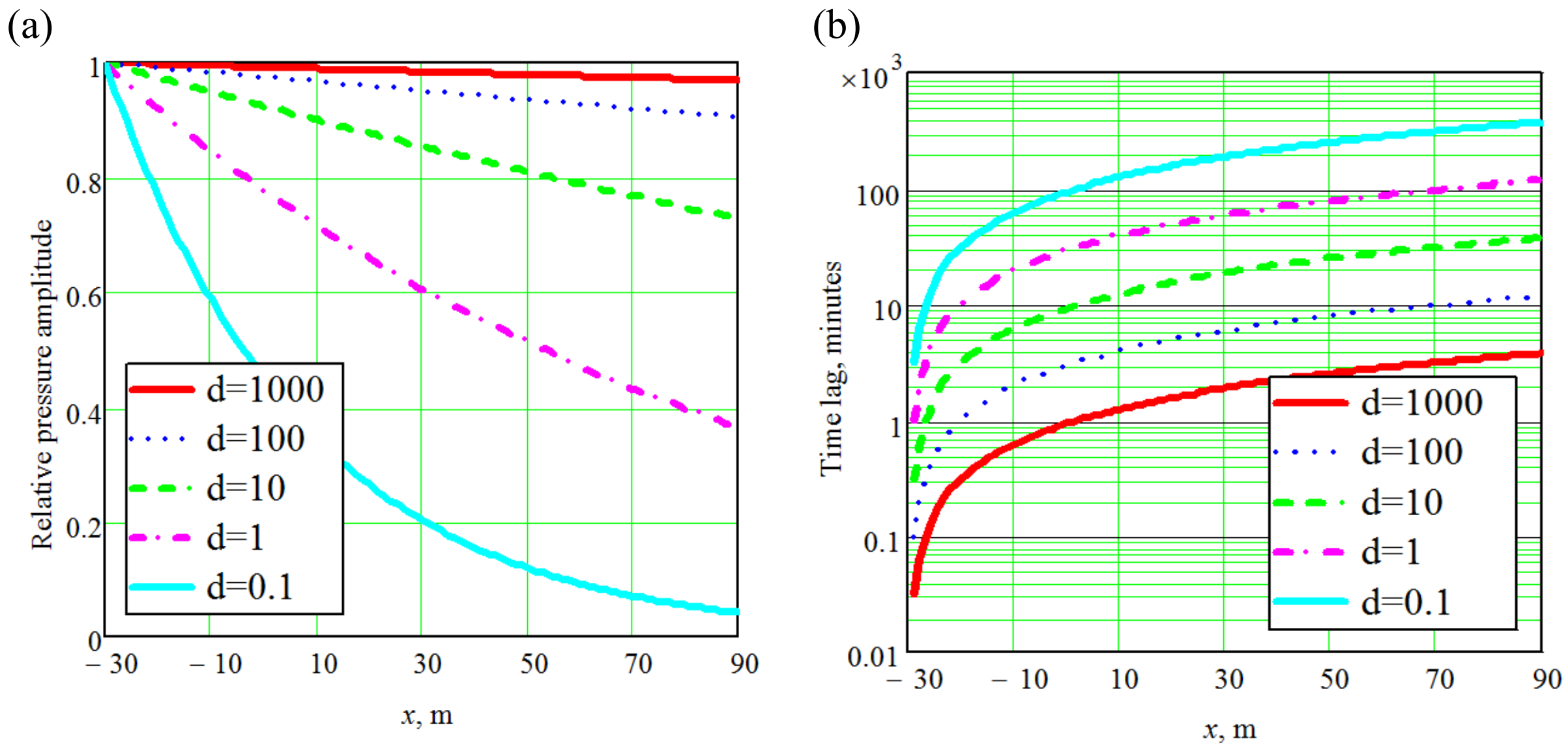

Figure 14. Analytical solution of 1D problem of tidal oscillations in semi-infinite saturated aquifer. (a) - relative pressure amplitude distribution along the domain; (b) - time lag, minutes, along the domain – in logarithmic scale

From (14) follows that:

$$C = \frac{A}{2i\sinh(\sqrt{i\frac{\omega}{d}}L)}, \tag{15}$$

(13)+(15) $\Rightarrow$

$$p(x,t)\Big| = 2\,\mathrm{Re}\left\{ \frac{A\cdot\sinh(\sqrt{i\frac{\omega}{d}}x)}{2i\cdot\sinh(\sqrt{i\frac{\omega}{d}}L)}(\cos(\omega t) + i\cdot\sin(\omega t)) \right\}, \tag{16}$$

Taking into account that

$$\sinh(\sqrt{\frac{i\omega}{d}}L) = \cos(\sqrt{\frac{\omega}{4d}}L)\sinh(\sqrt{\frac{\omega}{4d}}L) + i\cdot\sin(\sqrt{\frac{\omega}{4d}}L)\cosh(\sqrt{\frac{\omega}{4d}}L), \tag{17}$$

(16) can then be written as:

$$\begin{aligned} &p(x,d,L,t) = p_A(x)\cdot\sin(\omega(t-\Delta t)),\\ &p_A(x,d,L) = A\sqrt{\frac{\cosh(x\sqrt{\frac{\omega}{d}}) - \cos(x\sqrt{\frac{\omega}{d}})}{\cosh(L\sqrt{\frac{\omega}{d}}) - \cos(L\sqrt{\frac{\omega}{d}})}},\\ &\Delta t(x,d,L) = \begin{cases} \frac{1}{\omega}\cdot arctg\left[\frac{\exp r1}{\exp r2}\right] & if\ \ \exp r2 > 0 \\ \frac{1}{\omega}(\pi + arctg\left[\frac{\exp r1}{\exp r2}\right])\ otherwise \end{cases}, \end{aligned} \tag{18}$$

where (19)

$$\exp r1 = -\sinh(\sqrt{\frac{\omega}{4d}}(x+L))\cdot\sin(\sqrt{\frac{\omega}{4d}}(x-L)) + \sinh(\sqrt{\frac{\omega}{4d}}(x-L))\cdot\sin(\sqrt{\frac{\omega}{4d}}(x+L)),$$

$$\exp r2 = \cosh(\sqrt{\frac{\omega}{4d}}(x+L))\cdot\cos(\sqrt{\frac{\omega}{4d}}(x-L)) - \cosh(\sqrt{\frac{\omega}{4d}}(x-L))\cdot\cos(\sqrt{\frac{\omega}{4d}}(x+L))$$

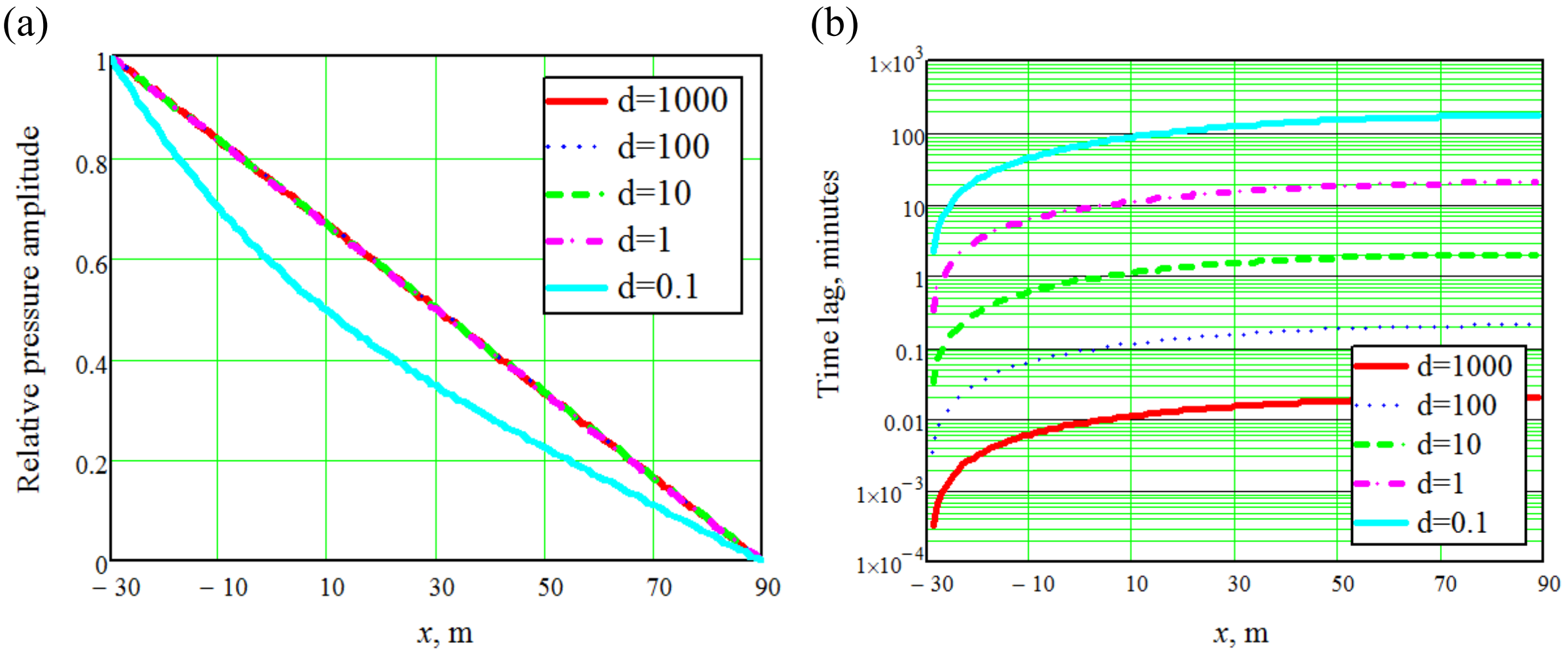

Figure 15. Analytical solution of 1D problem of tidal oscillations in bounded saturated aquifer. (a) - relative pressure amplitude distribution along the domain; (b) - time lag, minutes, along the domain – in logarithmic scale

For dense soils with diffusivity $d$<=1 m$^2$/s, the analytical model predicts non-linear profiles of pressure distribution, however the absolute values of pore pressure do not agree with 2D numerical simulation. For example, for d=1 m$^2$/s, analytical relative pressure amplitude in point x=50 m $P_A$ = 0.511, while in 2D numerical solution simulated amplitude $P_A$= 0.2.
Possible sources of mismatch between 1D analytical and 2D numerical models are two-dimensional flow behavior at the sea-side and diffusion of water above the phreatic line, which is considered in 2D numerical model only.

Calibration of the LiveDike soil parameters based on the sensitivity analysis is described in Section 7.

## 7 Calibration of diffusivities for the Livedike

Calibration has been performed for the first section of the dike. As it was mentioned in section 5, we have to find the values of 6 parameters: lengths of homogeneous zones $L_1$, $L_2$ and diffusivities $d_1$, $d_2$, $d_3$, $d_4$ (see Figure 13). Below we describe the procedure of diffusivity calibration using measured data from 3 sensors: E3, E4 and G2 in Figure 13. The algorithm is generic and can be used for any number of sensors in a dike cross-section.

Initial values of $L$ and $d$ parameters are obtained by superposition of analytical solutions derived from solution (18) for various periodic boundary conditions:

- In the 1st zone ($d$=$d_1$, 0<$x_1$<$L_1$): $p_1(x_1,t) = p_{11}(x_1,t) + p_{12}(x_1,t)$, where $p_{11}(x_1,t)$ is a solution of (8) with the boundary conditions:

$$p_{11}(x_1,t)\big|_{x_1=0} = A\sin(\omega t),\ p_{11}(x_1,t)\big|_{x_1=L_1} = 0\,, \qquad (20)$$

  and $p_{12}(x_1,t)$ is a solution of (8) with the boundary conditions:

$$p_{12}(x_1,t)\big|_{x_1=0} = 0, p_{12}(x_1,t)\big|_{x_1=L_1} = A_{interface1}\sin(\omega t + \varphi_{interface1}), \quad (21)$$

Here $A$ is tidal amplitude, $\omega$ is tidal frequency, $A_{interface1}$, $\varphi_{interface1}$ are local amplitude and phase delay on the interface of zones #1 and #2 (not known a priory, to be determined from a continuity condition (23);

- In the 2[d] zone ($d$=$d_2$, 0<$x_2$<$L_2$): $p_2(x_2,t)$ is a solution of (8) with the boundary conditions:

$$p_2(x_2,t)\big|_{x_2=0} = A_{interface1}\sin(\omega t + \varphi_{inetrface1}), p_2(x_2,t)\big|_{x_2=L_2} = 0;$$

- Continuity condition for the interface between 1[st] and 2d zones states that the value of flow velocity does not change at the interface:

$$\frac{\partial}{\partial x_1} p_1(x_1,t)\bigg|_{x_1=L_1} = \frac{\partial}{\partial x_2} p_2(x_2,t)\bigg|_{x_2=0} \quad (23)$$

From equation (20), we obtain two independent conditions: one for oscillation amplitude $A_{interface1}$ and one for oscillation phase $\varphi_{interface1}$.

- In the 3[d] zone: ($d$=$d_3$, 0<$x_1$<$L_1$): $p_3(x_1,t) = p_{31}(x_1,t) + p_{32}(x_1,t)$, where $p_{31}(x_1,t)$ is asolution of (8) with the boundary conditions:

$$p_{31}(x_1,t)\big|_{x_1=0} = A\sin(\omega t), p_{31}(x_1,t)\big|_{x_1=L_1} = 0 \quad (24)$$

$p_{32}(x_1,t)$ is a solution of (8) with the boundary conditions:

$$p_{32}(x_1,t)\big|_{x_1=0} = 0, p_{32}(x_1,t)\big|_{x_1=L_1} = A_{interface2}\sin(\omega t + \varphi_{interface2}), \quad (25)$$

where $A_{interface2}$, $\varphi_{interface2}$ are unknown local amplitude and phase delay on the interface of zones #3 and #4;

- In the 4[th] zone ($d$=$d_4$, 0<$x_2$<$L_2$): $p_4(x_2,t)$ is a solution of (8) with the boundary conditions:

$$p_4(x_2,t)\big|_{x_2=0} = A_{interface2}\sin(\omega t + \varphi_{inetrface2}), p_4(x_2,t)\big|_{x_2=L_2} = 0 \quad (26)$$

- Continuity condition for the interface between 3[d] and 4[th] zones is:

$$\frac{\partial}{\partial x_1} p_3(x_1,t)\bigg|_{x_1=L_1} = \frac{\partial}{\partial x_2} p_4(x_2,t)\bigg|_{x_2=0} \quad (27)$$

Similar to (23), (27) gives 2 scalar conditions: one for oscillation amplitude and one for oscillation phase

Equations (23, 27) together with 6 conditions equating amplitudes and time lags in virtual sensors with those in real sensors E3, E4, G2 form a system of 10 scalar equations to determine initial guess values for the parameters $L_1$, $L_2$, $d_1$, $d_2$, $d_3$, $d_4$, $A_{interface1}$, $\varphi_{interface1}$, $A_{interface2}$, $\varphi_{interface2}$.

To find more accurate values of $d_1$, $d_2$, $d_3$, $d_4$ we run numerical simulations as described in section 4, compare the results with real sensor data and tune the parameters. For a training period of 48 hours, the following parameters values have been obtained:

| (Horizontal diffusivity)·(water viscosity), Pa·m$^2$ | | | | Zone lengths | |
|---|---|---|---|---|---|
| $d_1 \cdot \mu$ | $d_2 \cdot \mu$ | $d_3 \cdot \mu$ | $d_4 \cdot \mu$ | $L_1$, m | $L_2$, m |
| $0.1 \cdot 10^{-3}$ | $0.01 \cdot 10^{-3}$ | $0.9 \cdot 10^{-3}$ | $0.01 \cdot 10^{-3}$ | 82 | 13 |

Simulation results for "training" period are shown in Figure 16(a), for the E4 pressure sensor.

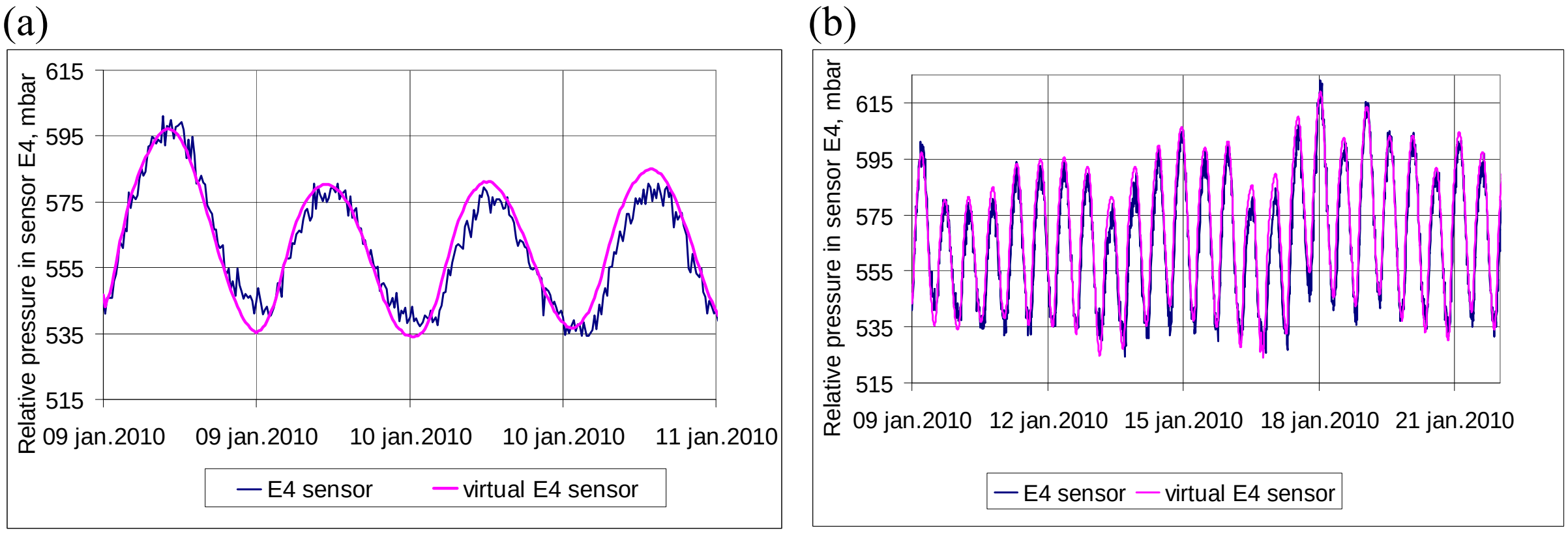

Figure 16. Relative pore pressure dynamics in sensor 1E4 with calibrated soil properties: (a) Comparison of real sensor data (blue) with simulation results (magenta) on a training data set; (b) The same, for a longer period of 12 days

For a long-term behavior, slow attenuated dynamics of ground water level $h_{gw}(t)$ at the land side of the dike should be represented in the boundary condition. Long-term simulations for January 2010 and August 2009 periods have been performed. The attenuated signal $h_{gw}(t)$ has been obtained by averaging the tidal signal $h(t)$ with a one-day sliding window and multiplying it by dissipation coefficient *q:* $h_{gw}(t) = q \cdot h(t)_{averaged}$ (see Figure 17 for January ground water table). The value of $q$ varied depending on the season ($q$=0.15 for January and $q$=0.25 for August). Looking at Figure 17, we can see that the averaged tidal signal represents slow oscillations with period varying between 2 and 3 days. Variation of dissipation coefficient $q$ with the season qualitatively agrees with analytical solution (12) for propagation of slow fluctuations in homogeneous aquifer: according to (12), $q = e^{-x\sqrt{\frac{\pi}{Td}}} \Rightarrow$

$\Rightarrow q_{august} = e^{-(L_1+L_2)\sqrt{\frac{\pi}{Td}}} = 0.28$, where homogeneous aquifer diffusivity $d$=0.1 m$^2$/s, for slow oscillations with period $T$=48 hrs, aquifer length $L_1$+$L_2$=95 m;

$\Rightarrow q_{january} = e^{-(L_1+L_2)\sqrt{\frac{\pi}{Td}}} = 0.18$, for the aquifer with diffusivity $d$=0.1/1.8 m$^2$/s (which is summer diffusivity scaled by $\mu_{january} / \mu_{august}$), $T$=48 hrs, $L_1$+$L_2$=95 m.

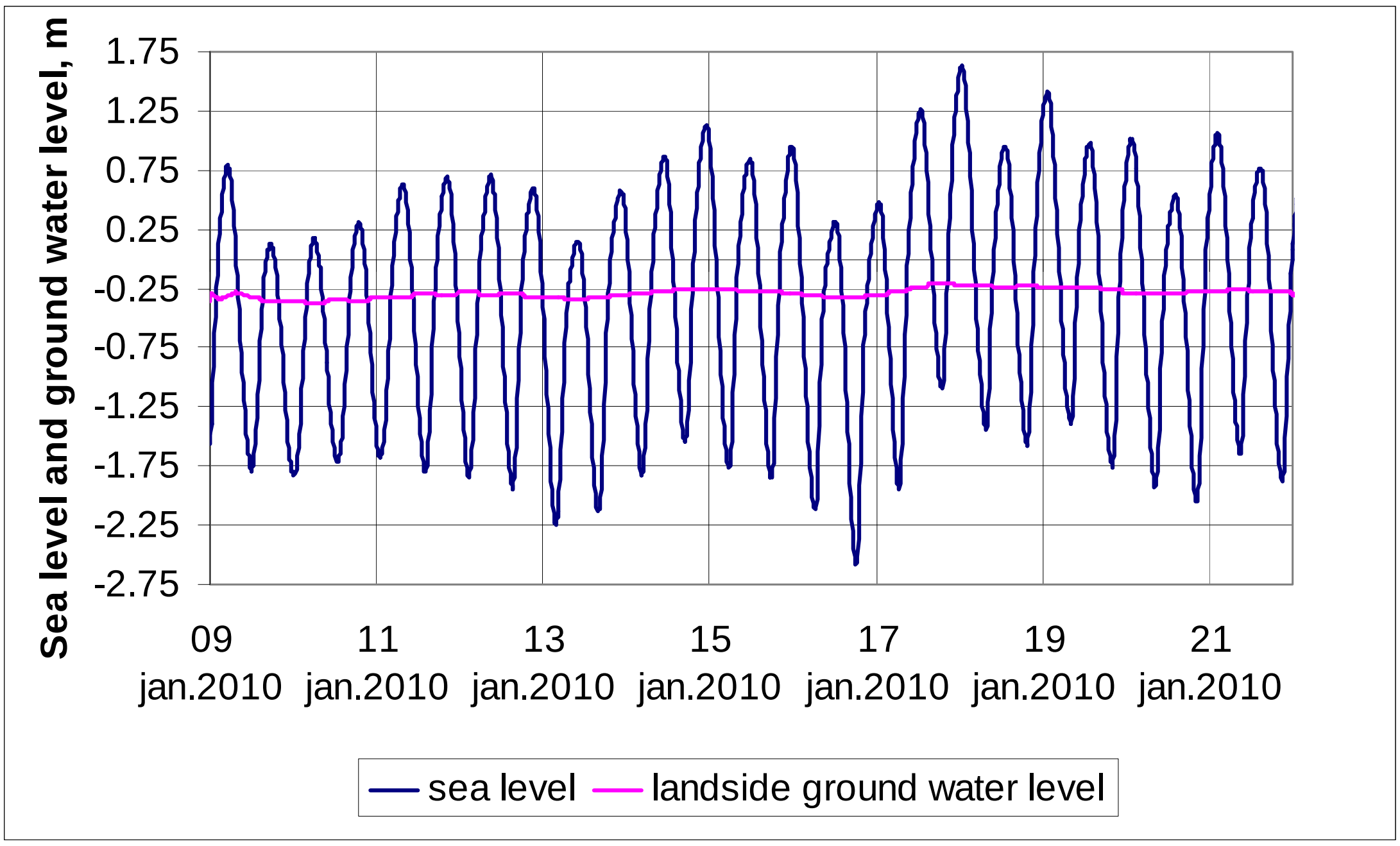

Figure 17. Livedike calibration. Sea level $h(t)$ and attenuated ground water level $h_{gw}(t)$ at the land side boundary

## 8 Conclusions

A finite element solver for analysis of earthen dikes stability has been developed and integrated into the UrbanFlood early warning system for early flood protection, where the simulation can be run with real-time input from water level sensors or with predicted high water levels due to upcoming storm surge or river flood. In the first case, comparison of simulated pore pressures with real data can indicate a change in soil properties or in dike operational conditions (e.g. failure of a drainage pump). In the second case, simulation can predict the structural stability of the dike and indicate the "weak" spots in the dikes that require attention of dike managers and city authorities.

Mathematical and finite element models of earthen dike behavior under dynamic hydraulic load have been developed. Transient flow through porous media was modeled by Richards equation with van Genuchten model for water retention in partially saturated zone above the phreatic surface.

Sensitivity analysis of porous flow dynamics to soil diffusivity showed that:

1. Distribution of pore pressure amplitudes across the dike (in horizontal direction from the sea) is close to linear for highly permeable soils (like gravels and coarse sands) and is significantly non-linear for non-permeable soils, such as clays.
2. Pressure amplitude for coarse media ($d \geq 10$ m$^2$/s) is insensitive to the value of diffusivity, and is only defined by boundary conditions.
3. The time delay is always sensitive to the value of diffusivity and can be calibrated by choosing appropriate saturated diffusivity to match sensor data.

A generic procedure for calibration of diffusivities in heterogeneous dike has been proposed and successfully tested on the first cross-section of the Livedike (Groningen). Calibration has been performed on tidal data sets obtained from real pore pressure sensors. Simulation results with calibrated soil parameters match experimental data, not only on the "training set" but also for a much longer period of time.

One-dimensional harmonic solution discussed above qualitatively agrees with 2D numerical solution; analytical solutions are employed in the numerical model to get initial guess values for diffusivities in the heterogeneous soil build-up.

Our future plans include implementing the program for automatic diffusivities calibration, with the algorithm based on the generic calibration procedure described above.

## Acknowledgements

This work is supported by the EU FP7 project *UrbanFlood*, grant N 248767; by the Leading Scientist Program of the Russian Federation, contract 11.G34.31.0019; and by the BiG Grid project BG-020-10, # 2010/01550/NCF with financial support from the Netherlands Organisation for Scientific Research NWO. It is carried out in collaboration with AlertSolutions, Deltares, IJkDijk Association, Rijkswaterstaat, SARA Computing and Networking Services, Waterschap Noorderzijlvest.